\RequirePackage[hyphens]{url}
\documentclass[a4paper,11pt]{article}
\usepackage[bindingoffset=0.2in,
            left=1in,
            right=1in,
            top=1in,
            bottom=1in,
            footskip=.25in]{geometry}
\newcommand*\patchAmsMathEnvironmentForLineno[1]{%
\expandafter\let\csname old#1\expandafter\endcsname\csname #1\endcsname
\expandafter\let\csname oldend#1\expandafter\endcsname\csname end#1\endcsname
\renewenvironment{#1}%
{\linenomath\csname old#1\endcsname}%
{\csname oldend#1\endcsname\endlinenomath}}%
\newcommand*\patchBothAmsMathEnvironmentsForLineno[1]{%
\patchAmsMathEnvironmentForLineno{#1}%
\patchAmsMathEnvironmentForLineno{#1*}}%
\AtBeginDocument{%
\patchBothAmsMathEnvironmentsForLineno{equation}%
\patchBothAmsMathEnvironmentsForLineno{align}%
\patchBothAmsMathEnvironmentsForLineno{flalign}%
\patchBothAmsMathEnvironmentsForLineno{alignat}%
\patchBothAmsMathEnvironmentsForLineno{gather}%
\patchBothAmsMathEnvironmentsForLineno{multline}%
}

\usepackage[doublespacing]{setspace}
\usepackage[export]{adjustbox}
\usepackage[hidelinks]{hyperref}
\usepackage{breakurl}
\usepackage{graphicx}
\usepackage{amssymb}
\usepackage{natbib}
\usepackage{enumitem}
\usepackage{physics}
\usepackage{graphicx}
\usepackage{placeins}
\usepackage{array}
\usepackage{xcolor}
\usepackage[normalem]{ulem}
\usepackage{endfloat}
\usepackage[displaymath, mathlines]{lineno}
\setlist{topsep=0.75em, itemsep=0.75em}
    \makeatletter
\def\@fnsymbol#1{\ensuremath{\ifcase#1\or \mathsection\or \dagger\or
   \ddagger\or \mathparagraph\or \|\or **\or \dagger\dagger
   \or \ddagger\ddagger \else\@ctrerr\fi}}
    \makeatother
\title{A decade of airborne electromagnetic surveying Lake Menindee (Australia) under varying water levels}
\author{A. Ray$^\dagger$, A. McPherson$^\dagger$, R. C. Brodie${^\dagger}{^\dagger}$, A. Y. Ley-Cooper$^\dagger$, K. P. Tan$^\dagger$, M. Hatch$^{\ddagger,*}$,\\R. N. Deo$^\dagger$, S. Wong$^\dagger$, F. Dauti$^{**}$, W. Cook$^\dagger$, T. Scarr$^\dagger$ and Y. Sun$^\Delta$}
\date{\footnotesize
$^\dagger$Geoscience Australia\thanks{email: anandaroop.ray@ga.gov.au},\\
${^\dagger}{^\dagger}$ formerly at Geoscience Australia, \\ 
$^\ddagger$University of Adelaide, Australia,\\
$^*$VistaClara Ltd.,\\
$^{**}$ University of Milan, Italy,\\
$^\Delta$ National Computational Infrastructure, Australia\\
}                                           
\newcommand{\Var}{\mathrm{Var}}
\newcommand{\Cov}{\mathrm{Cov}}
\newcommand{\E}{\mathrm{\textbf{E}}}
\begin{document}
\maketitle
\begin{abstract}
Time domain airborne electromagnetic (AEM) surveying is a mature geophysical tool for imaging the Earth's shallow subsurface. It produces images of the electromagnetic conductivity structure of the earth, down to depths of a few hundred metres. The AEM method is fast, with rotary-wing or fixed-wing aircraft acquiring data at speeds of 100-300 km/hr, making it an ideal near-surface reconnaissance tool. The physics of the AEM method are sensitive primarily to the subsurface conductivity, which is influenced by a range of geological factors such as mineral content, porosity, and water content and chemistry. In addition, the inferred subsurface conductivity depends on the accurate measurement and modelling of airborne transmitter and receiver geometries -- a challenging task given the speed of acquisition and variability of wind conditions during an acquisition flight. In this work, we present inferences of the subsurface conductivity over Lake Menindee, New South Wales, Australia, using data from test flights and various AEM systems over a ten year period (2014-2024). The lake storage has varied dramatically over this time, and the test flights have coincided with both high and low water levels. While this difference in storage volume undoubtedly influences the near surface conductivity, a remarkably consistent interpretation of the regional geology is possible regardless of the hydrologic conditions. While the upper ten metres of the modelled depth sections exhibit the greatest time-variability in inferred electromagnetic conductivity, {the hypothesis that lakebed near-surface conductivity is significantly correlated with the lake water volume cannot robustly be established}. We also provide some information theoretic calculations for each inversion result to aid in their quantitative comparison. The implications of our study are that subtle, shallow, hydrogeological changes are difficult to image with repeat AEM overflights from different systems. Conversely, we establish that different AEM systems with minimal extra processing robustly image the regional geo-electric structure of the near surface, validated by known stratigraphy and associated geological information, as well as borehole conductivity logs. 
\end{abstract}
\section{Introduction}
The value of airborne electromagnetic (AEM) surveying is partly dependent on the sensitivity of the acquired data to subsurface geological features of interest, calibration of the physics involved, and repeatability of the AEM system that is used. These factors are important in assessing the potential for a particular AEM system to meet the aims of a survey. AEM systems have evolved in their use from ``bump-finding" anomalous mineral accumulations \citep[e.g.,][]{Gilgallon2019} to regional stratigraphic mapping and groundwater assessment \citep[e.g.,][]{Ley-Cooper2020, Macca2024}. Inversion of the transient AEM data for subsurface conductivity directly leads to geo-electric imaging, obviating the need for any further ``processing''. In essence, it is a full-waveform technique that relies on an accurate description of the physics involved, transmitter-receiver separations and rotation angles, as well as noise in the observations. Once the AEM acquisition system, noise levels and inversion parameters are calibrated, no further calibration of the derived subsurface conductivities is required. Consequently, inverted conductivities from different surveys are comparable. 

For testing different AEM systems, it is therefore imperative to carry out quantitative inversion of acquired AEM transients, over areas of reasonably known geological structure and subsurface conductivity. Furthermore, while quantitative inversions of AEM transients can be robustly and effectively carried out in a deterministic fashion, the nonlinear, ill-posed and non-unique nature of the AEM inverse problem necessitates the use of regularisation techniques to invert a stable, interpretable model \citep[e.g.,][]{Constable1987, Farquharson1998, Auken2004}. The use of regularisation however, has inherent in it, certain disadvantages. For instance, some regularisation schemes force a return to a background ``prejudice'' model \citep[e.g.,][]{Key2016} when sensitivity to the data are lost, and the smoothest possible model to fit the data, though a wise interpretational choice, is an extremal model.
Probabilistic or stochastic inversions, when carried out with appropriate prior information, provide an ensemble of posterior models that are compatible with both the prior information, as well as the observed data and its noise statistics \citep[e.g.,][]{Scales1997, tarantolainvbook}. This in turn reduces the uncertainty of geological interpretation, and makes it easier to compare AEM systems over the same stretch of land. 
\section{{Study area and available data}}
\begin{figure}
	\centering
	\adjincludegraphics[height=0.65\height,trim={0 {.0\height} {.55\width} {.0\height}},clip]{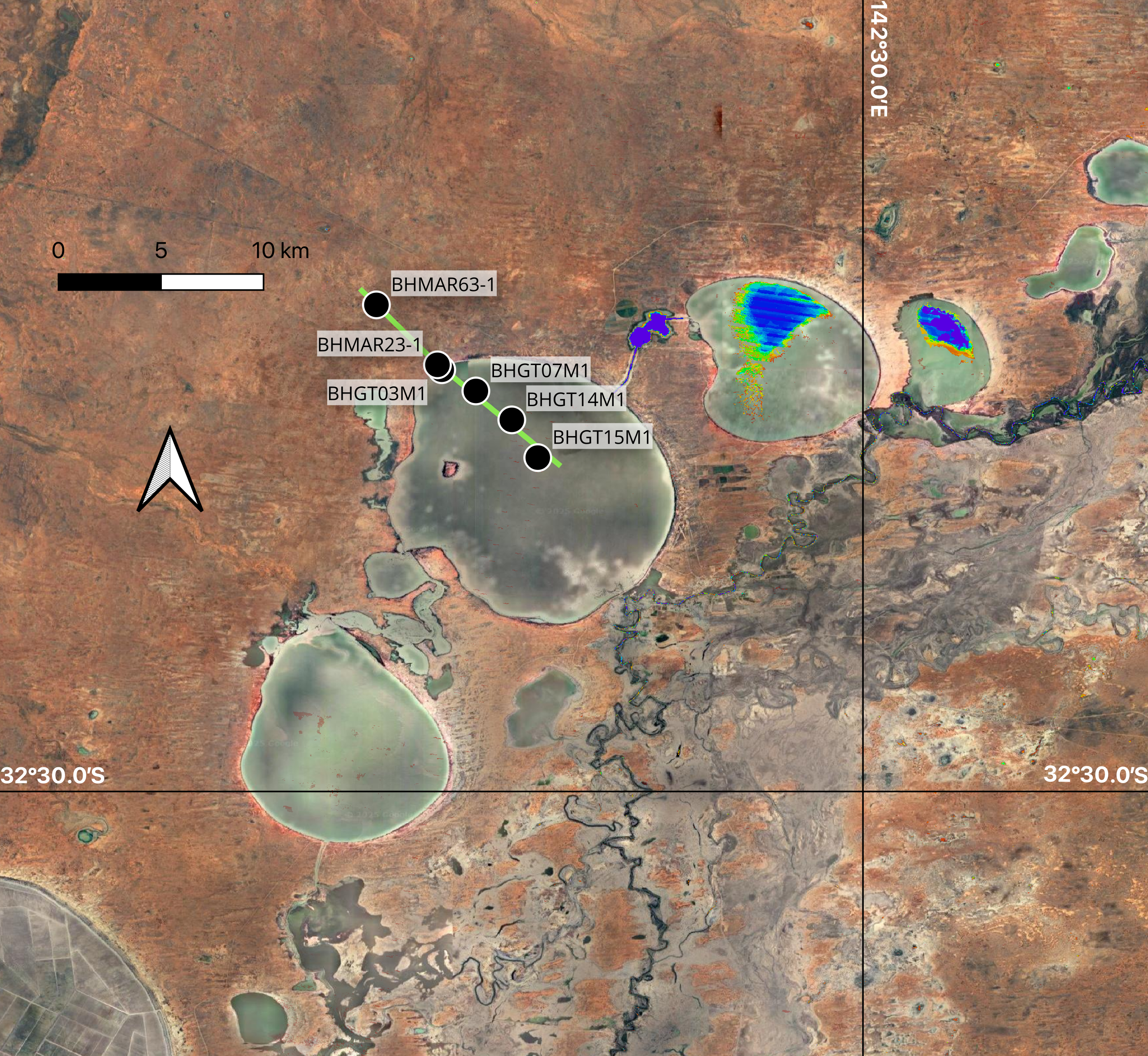}
	\hspace{-6.0pt}
	\adjincludegraphics[height=0.65\height,trim={{.45\width} {.0\height} 0 {.0\height}},clip]{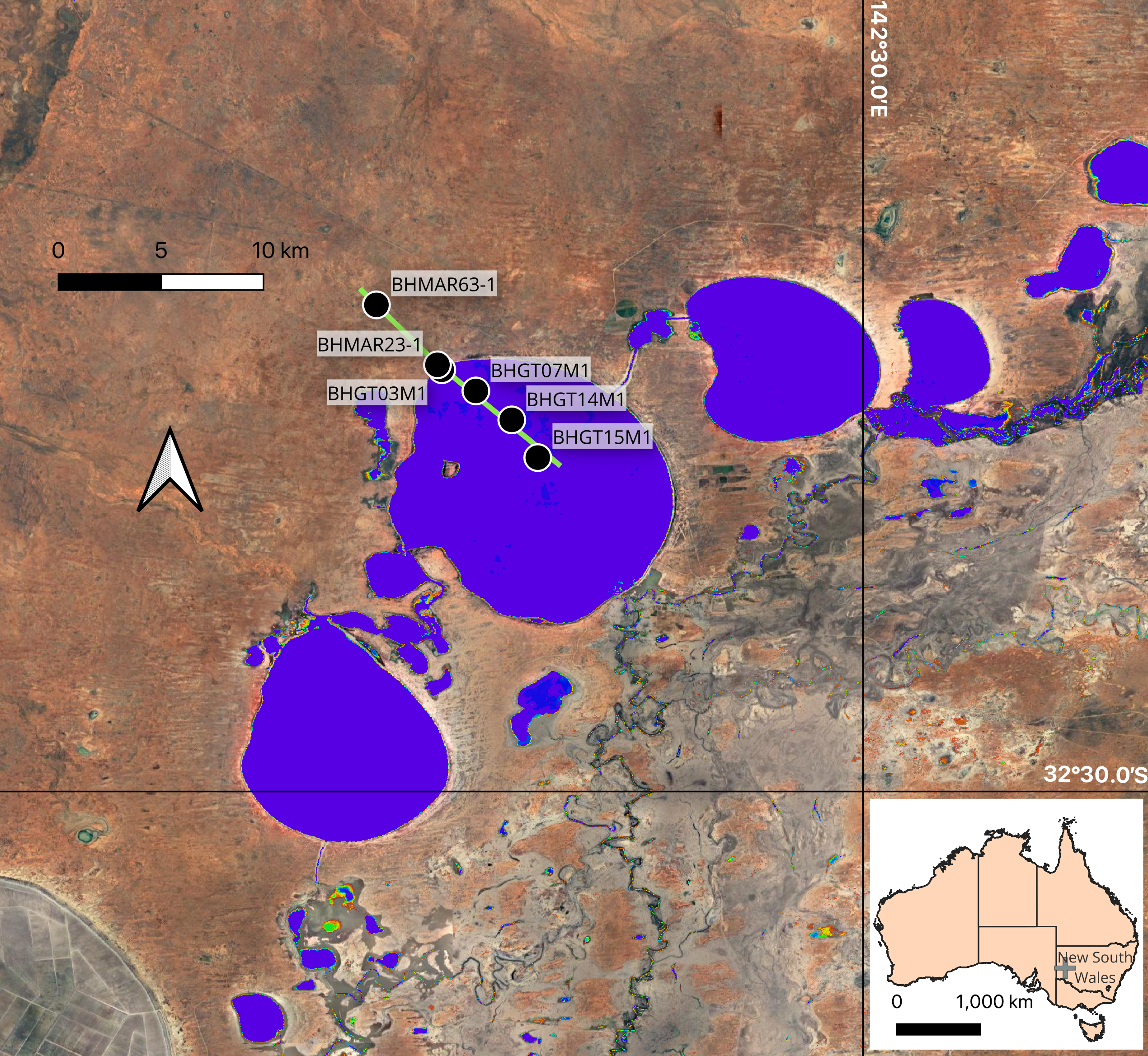}
	\includegraphics[width=0.925\linewidth]{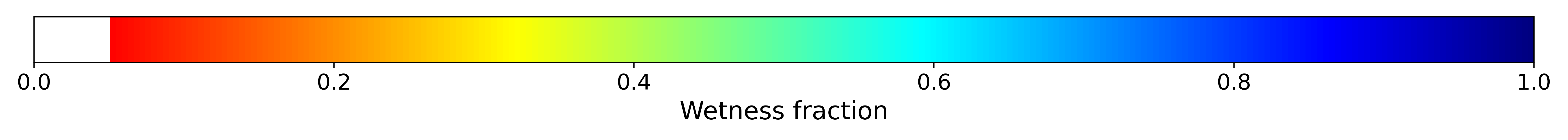}
	\caption{The repeat line section (green line) of the Menindee AEM test range in New South Wales, Australia. This section of the test range is proximal to various boreholes, and Lake Menindee itself has been subject to 7 m changes in water level over time. On the right we show Google surface imagery, overlain by summary April-October Landsat water observations from space \citep{Mueller2015} in 2023, when the lake is completely full. On the left we show summary water observations for the same period in 2019, prior to the Black Summer bushfires in Australia -- the lake bed is completely dry. In both 2019 and 2023 in areas without appreciable wetness, the wetness fraction is transparent, showing the underlying Google surface imagery. The inset map shows the location of the test range (grey cross) within the Australian state of New South Wales.}
	\label{pic:map}
\end{figure}

As a means of assessing data acquired by different AEM systems, Geoscience Australia established the Menindee Airborne Electromagnetic test range in 2017, near the town of Menindee, approximately 100 km southeast of Broken Hill in western New South Wales, Australia (Figure~\ref{pic:map}). The site was chosen to take advantage of considerable prior knowledge collected by Geoscience Australia {during the Millenium Drought \citep[see e.g.,][]{https://doi.org/10.1002/wrcr.20123}} including downhole geophysical and stratigraphic logging, geological interpretation, and a LiDAR elevation survey. 

While a rigorous probabilistic comparison of the subsurface information content in a fixed-wing AEM system (TEMPEST) and a helicopter AEM system (SkyTEM 312) over Lake Menindee can be found in \cite{Ray2023}, in the work presented here we add to the comparisons, six more surveys from both earlier and later than previously considered {(Table~\ref{tab:info})}. These surveys involve the addition of another fixed-wing system (SPECTREM), and four more rotary-wing systems (VTEM, Xcite, SkyTEM-312HP-LF, HeliTEM-LF). These AEM systems were not chosen with a particular subsurface target in mind, but were available for assessment in order to qualify for Geoscience Australia's {continent-wide} survey tender process. 
\begin{table}
\centering
\begin{tabular}{|>{\centering}p{0.14\textwidth}|>{\centering}p{0.11\textwidth}|>{\centering}p{0.12\textwidth}|>{\centering}p{0.1\textwidth}|>{\centering}p{0.25\textwidth}|c|}
\hline
\textbf{System} &
\textbf{Aircraft} &
\textbf{Waveform} &
\textbf{Base freq. in Hz} &
\textbf{Flight geometry nuisances inverted} &
\textbf{Year} \\
\hline
VTEM & Helicopter & Pulse & 25 & None & 2014 \\
\hline
SkyTEM-312 & Helicopter & Dual pulse & 25 & None & 2015 \\
\hline
TEMPEST & Fixed-wing & Half square & 25 & Tx-Rx inline separation, Rx height & 2017 \\
\hline
SPECTREM & Fixed-wing & Square & 25 &
Tx-Rx inline separation, Rx height & 2018 \\
\hline
Xcite & Helicopter & Pulse & 25 & None & 2019 \\
\hline
HeliTEM LF & Helicopter & Pulse & 6.25 & None & 2023 \\
\hline
SkyTEM-312 LF & Helicopter & Dual pulse & 12.5 & None & 2024 \\
\hline
Xcite & Helicopter & Pulse & 25 & None & 2024 \\
\hline
\end{tabular}
\caption{{A description of the eight different systems flown over the same line in Figure~\ref{pic:map}, together with the modelling nuisance parameters that were inverted. By `pulse' we mean a waveform which can be approximated effectively with a hundred linear ramps or less. The half square and square waveforms have a duty-cycle of 50\% and 100\% respectively, however they are intrinsically `pulse waveforms' for which the data are processed and presented as `equivalent square waveform' magnetic field data. Tx and Rx here refer to transmitter and receiver, respectively. The consideration of Tx-Rx nuisance parameters in the inversion is explained in Sections~\ref{sec:methods} and~\ref{sec:consider}. Flight years are shown in the last column.}}
\label{tab:info}
\end{table}
\section{{Methods}}\label{sec:methods}
\begin{figure}
	\centering
	\includegraphics[width=\linewidth]{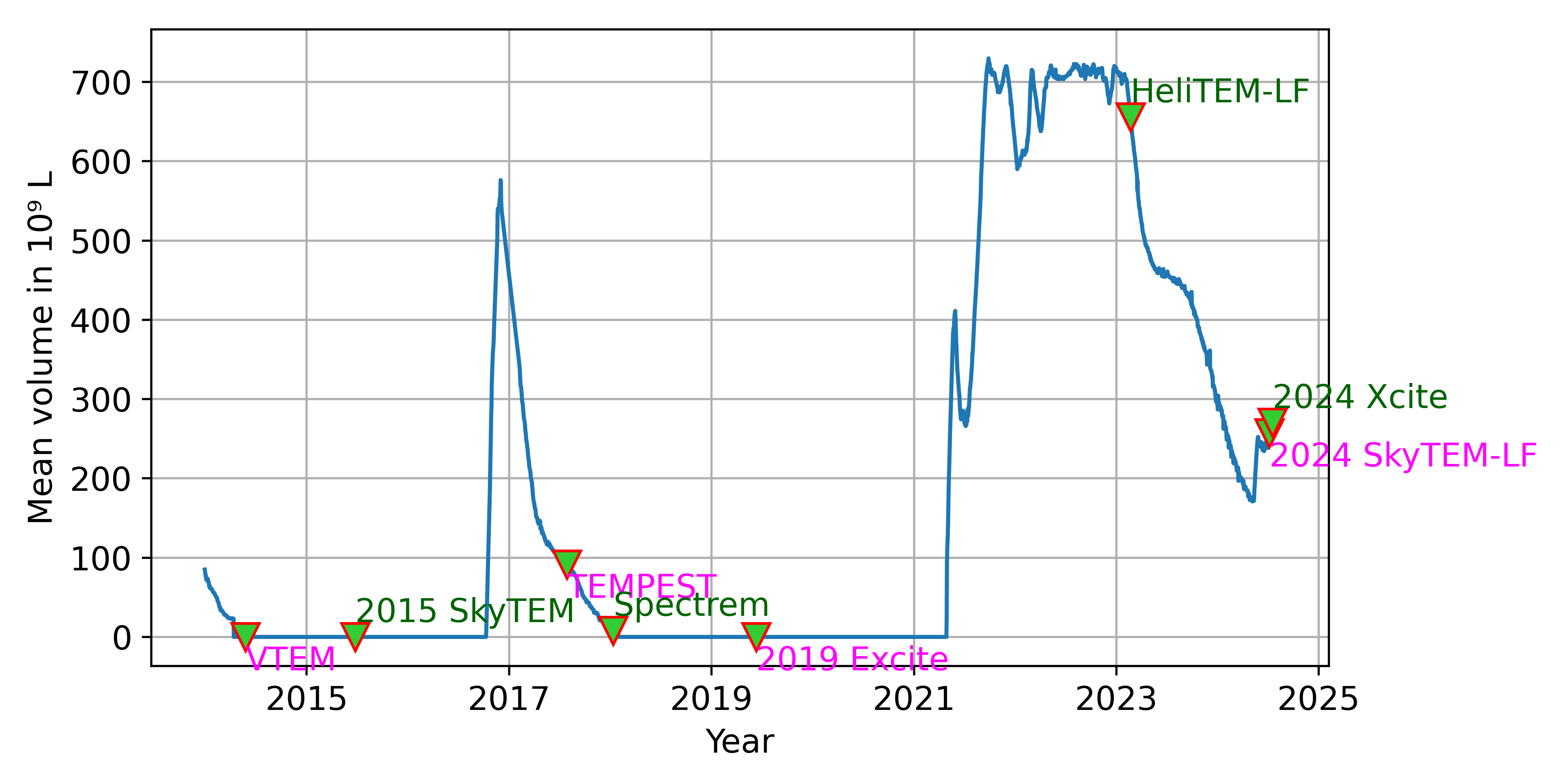}
	\caption{Time series of water volume in Lake Menindee reservoir. The triangular markers labelled by acquisition system indicate the water levels during 8 test flights over 10 years. The lake is full when water volume is over 600 Gigalitres, e.g., during the 2023 HeliTEM overflight. Source: \url{https://waterinsights.waternsw.com.au/}}
	\label{pic:waterlevels}
\end{figure}

{We carry out our comparisons using both deterministic and probabilistic inversions, however, there have been 7 m variations in the Menindee reservoir water level and changes in water volume between 2014-2024 (Figure~\ref{pic:waterlevels}). Therefore, we might reasonably expect that the near surface conductivity imaged by all AEM systems would show changes linked to variation in water content and salinity. Consequently, we attempt to establish a relationship between shallow inverted conductivities from these eight surveys and the reservoir water volume and establish its significance}. Two surveys with ostensibly the same system (Xcite) have flown over the lake in both dry and wet conditions, 5 years apart. 

While we would expect the top 10 m to show conductivity variations between 2014-2024, this depth also corresponds to relatively early times in the AEM transient, where any inaccuracies in the description of transmitter-receiver geometries and source waveform, will strongly manifest themselves in the inverted conductivities. To compensate for this, for helicopter systems the data noise levels are highest at these times, and in fixed-wing systems the transmitter-receiver geometries are themselves inverted \citep[using the primary inducing field, see][]{Brodie2010, Ray2023}, to prevent artefacts from propagating into the subsurface conductivity. 

{\section{Previous inter-system comparisons and our contribution}}
While other workers in the field have carried out comparisons between AEM systems \citep[e.g.,][]{Bedrosian2016, Minsley2021a}, the work presented here analyses a unique combination of geological and geophysical factors relevant to subsurface imaging. This includes known variation in near surface water content, well documented local geology, downhole induction log data, and a near-complete representation of modern-day time domain AEM systems (both rotary and fixed-wing) flown over the same test line in a 10-year period. We hope our findings will be of value in appropriately assessing the utility of the AEM surveying method for near surface imaging and subsequent geological interpretation.
\section{Modelling and inversion of AEM data}
In this work we utilise the open source HiQGA \citep{Ray2023} codebase. It uses the time domain forward computation described in \cite{Blatter2018}, the source waveform convolution in \cite{Fitterman1987}, frequency-domain transverse electric (TE) mode layered earth formulation of \cite{Loseth2007a} implemented in \cite{Ray:2012fk}, 201-pt Hankel transforms as described in \cite{Key:2012ku}, and transmitter frame to receiver frame active and passive rotations provided in \cite{Fitterman2004}. The codebase was used to derive deterministic as well as probabilistic inversion conductivities from the AEM data. Identical hyper-parameter (i.e., geometry nuisance) settings, priors and 52-layer model parameterisation were used across all inversions. 
\subsection{Determininistic inversion}
A fast Occam inversion \citep{Key2016} with within-bounds modifications was used for deterministic conductivities, where the following objective function was minimised: 
\begin{align}
	\phi (\mathbf m) &= \Bigg(\mathbf{||W(d - f(m))||}^2 + \lambda^2\Big[||\mathbf{Rm}||^2 + \beta^2\mathbf{||m - m_0||^2}\Big]\Bigg).\\
	\intertext{$\phi (\mathbf m)$ is conveniently separated into the following form with parts to do with the $\chi^2$ data misfit $\phi_{\chi^2}$ and the model regularisation norm $\phi_m$ written as:}
	\phi (\mathbf m) &= \phi_{\chi^2} (\mathbf W, \mathbf m) + \lambda^2\phi_m(\mathbf m, \mathbf m_0, \mathbf R, \beta^2),\\
	\intertext{where the popularly used data misfit term $\phi_d$, normalised by the number of data observations per sounding $n$  is given by:} 
	\phi_d (\mathbf W, \mathbf m) &= \frac 1 n \phi_{\chi^2} (\mathbf W, \mathbf m) = \frac 1 n \mathbf{||W(d - f(m))||}^2.
\end{align}
The nonlinear AEM forward operator is $\mathbf f$, and AEM data (voltage for helicopter systems and magnetic field for fixed-wing systems) are contained in $\mathbf d$. The data covariance matrix $\mathbf{C}_d$ is related to the data weights $\mathbf{W}$ such that $\mathbf{C}_d^{-1} = \mathbf{W}^t\mathbf{W}$, and conductivities are represented by the model $\mathbf m$. $\lambda^2$ is the strictly positive trade-off parameter, $\mathbf R$ is a first or second differences operator, and $\mathbf{m_0}$ is the reference or ``prejudice'' model.  According to the fast Occam formulation, $\lambda^2$ is found by sweeping through a decreasing range of positive values until the misfit $\mathbf{||W(d - f(m))||}^2$ from the model in the last nonlinear iteration decreases to 0.7 of its previous value \citep{Farquharson1993, Brodie2010}. In the second stage, when enough nonlinear iterations have been run such that misfit decreases beneath the assigned data noise, bracketing and root finding is used to find the largest $\lambda^2$ that provides a model at the assigned AEM data noise. This ensures that the smoothest model compatible with the observations and their noise level is attained. However, the choice of $\beta^2$, another positive quantity, is left to the practitioner. Large $\beta^2$ values ensure a quicker return to the reference model, while smaller values pay more attention to keeping the model smooth though the quantity $||\mathbf{Rm}||^2$. For fixed-wing systems, transmitter-receiver geometry nuisances are inverted within predefined bounds per non-linear iteration, in an alternate step with a BFGS (Broyden–Fletcher–Goldfarb–Shanno) scheme \citep[see][]{wright2006numerical}. 
\subsection{Probabilistic inversion}
For probabilistic inversion, a Bayesian approach was used:
\begin{equation}
	p(\mathbf{m|d}) \propto p(\mathbf{d|m}) \cdot p(\mathbf{m}),
\end{equation}
where $p(\mathbf{m})$ is the prior model probability, $p(\mathbf{d|m})$ is the model likelihood obtained from the model misfit and data noise statistics, and $ p(\mathbf{m|d})$ is the posterior probability inference we seek. The posterior model probability updates our prior knowledge of the subsurface in a manner proportionate to the likelihood of the misfit. We used the trans-dimensional Gaussian process Markov chain Monte Carlo (McMC) inversion scheme \citep{Ray2019, Ray2021} together with the modifications for marginalising fixed-wing geometry nuisances provided in \cite{Ray2023}. While no regularisation is required for trans-dimensional Bayesian inversion, its characteristic property of ``natural parsimony'' \citep[see][]{Malinverno2002, Bodin2009a, Dettmer2010a} in the number of Gaussian basis functions required to parameterise a model, and a weak depth-varying correlation length as described by \cite{Blatter2021} ensures that earth models do not overfit the AEM data. Identical uniform priors for all systems over conductivity, depth and geometry nuisances (for fixed-wing systems) are exactly as described in \cite{Ray2023} and not repeated here. {Production inversions \cite[e.g.][]{Ray2024AUSAEMPhase1, Scarr2025AusAEMNEQ} converge within 200,000 samples with negligible change in posterior statistics after 125,000 samples while discarding the first 50,000 samples as burn-in. For this study, to the extent possible given our computational allocation, we doubled the number of total and burnin samples to further ensure unbiased inference. Therefore, for all inversions presented here, the McMC burn-in fraction was set to 0.25 of 400,000 total samples per parallel McMC chain. An identical setup and details of the parallel tempering sampling algorithm used in this work can be found in \cite{Ray2023}.} The likelihood is given by the term:
\begin{align}
p(\mathbf{d|m}) &\propto \exp(- \frac1 2 \mathbf{||W(d - f(m))||^2}),\\
			&= \exp(-\frac1 2 \phi_{\chi^2}(\mathbf W, \mathbf m)),\\
			&= \exp\Big(-\frac1 2 n\phi_d (\mathbf W, \mathbf m)\Big),
\end{align}
noting that the probabilistic negative log likelihood and half the deterministic $\chi^2$ data misfit are one and the same, with the exception of unchanging additive constants for all models. Another popular misfit measure in EM geophysics, the root mean square or RMS data error is given by $\sqrt{\phi_d}$.The notation used above explicitly shows the dependence of both probabilistic and deterministic inversion on the weighting of the data, $\mathbf W$ (usually a diagonal matrix of inverse data standard deviations per time gate). The data errors in each gate are assigned by adding noise from two sources in quadrature \citep{Green2003}: one source is proportional to signal amplitude, and another is high altitude noise measured away from ground influence. 
\subsection{Considerations for comparing inverted conductivities from different surveys} \label{sec:consider}
We should keep the following in mind when comparing conductivity sections inverted from each survey. The two fixed wing surveys compared are TEMPEST and SPECTREM. However, the SPECTREM survey provides no information on the receiver rotation angles and the receiver was assumed to be perfectly in line with the transmitter frame. The fixed-wing inversions were of the total amplitude of the horizontal (X) and vertical (Z) components of the observed magnetic field in the receiver frame, thus obviating the need for receiver pitch estimation \citep{Ley-Cooper2020}. Therefore, for fixed wing systems, similar to \cite{Ray2023}, we invert for transmitter-receiver inline distance and receiver height in addition to subsurface conductivity. Inaccurate receiver yaw measurements however, will propagate into the inverted model. While we could potentially invert for more components of the receiver geometry, there are too many trade-offs between free parameters and conductivity to do this robustly. Secondly, the HeliTEM-LF survey used a transmitter base frequency of 6.25 Hz, and the SkyTEM-LF survey used a 12.5 Hz base frequency. All other systems used the standard 25 Hz base frequency for transmitted waveforms for synchronous rejection of 25 Hz power line noise in Australia. {This information is summarised in Table~\ref{tab:info}.}  

Finally, though the inverted conductivities are comparable, they are not theoretically equal, even if nothing were to have changed in the earth in the time between surveys. This is because no two transmitter waveforms are exactly the same (even when the same AEM system is used), flying heights and transmitter-receiver geometries are different, and the deterministic inversion ultimately produces a regularised estimate of the subsurface conductivity. Whether probabilistic or deterministic, we need to keep in mind that conductivity from an AEM survey is an indirect inference, not an observation of data. Broadly speaking, all surveys ought to produce a consistent picture of the regional geology. 
\subsection{Deterministic inversion results}
\begin{figure}
	\centering
	\includegraphics[width=\linewidth]{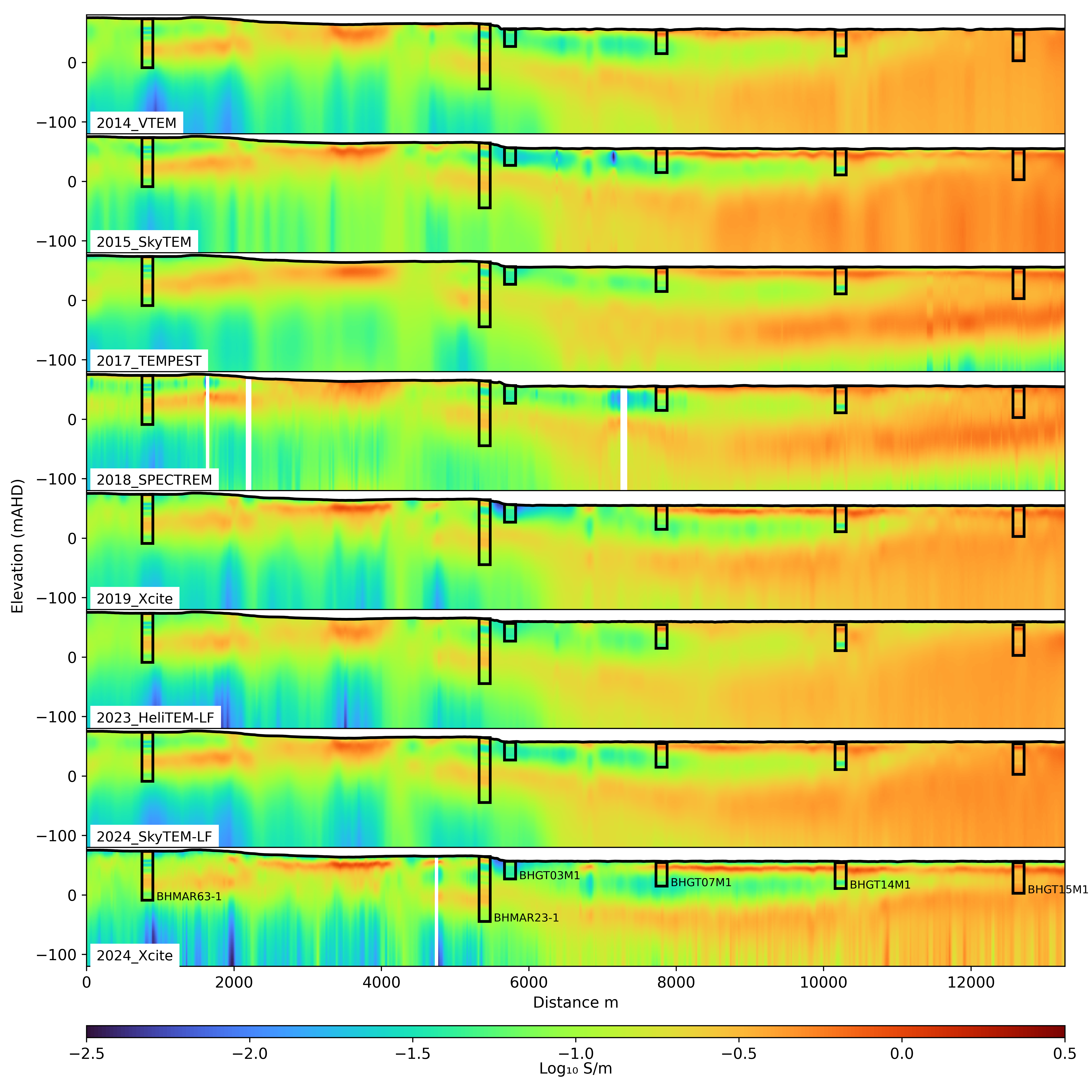}
	\caption{Conductivity-depth sections from a deterministic inversion of acquired AEM data, arranged chronologically from top to bottom. The northwest edge of the repeat line shown in Figure~\ref{pic:map} is to the left, and southeast to the right. Overlain induction log conductivies use the same colour scale as the AEM inversion. Warm colours are conductive and cool colours are resistive. Qualitatively, there is a close match between downhole conductivity values and the AEM derived conductivities, despite the intervening years and changing water volumes.}
	\label{pic:deterministic}
\end{figure}

\begin{figure}
	\centering
	\includegraphics[width=\linewidth]{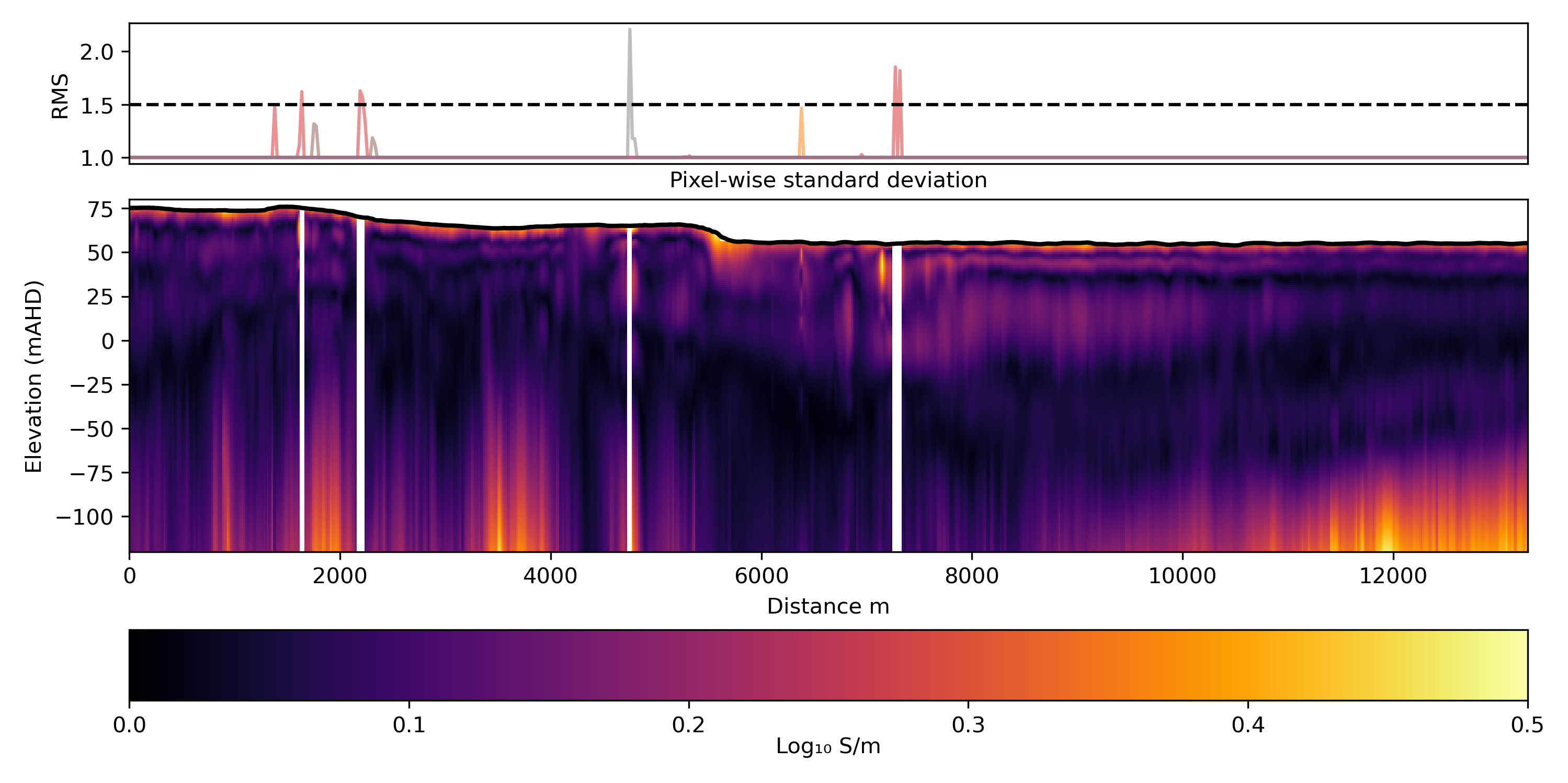}
	\caption{{Visualisation of the pixel-wise standard deviation between all the deterministically inverted conductivity sections seen in Figure~\ref{pic:deterministic}. The greater the variability, the higher the standard deviation (incandescent colours). Variability is greatest where inverted conductivity values are different, or have shifted between the different deterministic inversions. The top panel shows a plot of root mean square (RMS) misfit values from each inversion. At a few locations, the RMS values are higher than the threshold value of 1.5 (dashed line), where conductivities have been blanked out.}}
	\label{pic:deterministic_rms}
\end{figure}

All inversions were inverted with identical $\beta^2 = 0.0001$, ensuring only a weak adherence to the background conductivity $\mathbf m_0 = 0.01$ S/m, to a target RMS of 1.0. The comparison of inversions from each survey is shown in Figure~\ref{pic:deterministic}, with RMS data misfits higher than 1.5 blanked out. Also overlain are downhole induction log data in the same colour scheme as the inverted conductivities, and the qualitative agreement is remarkably striking. We should also note that the boreholes were logged between 2008-2010 towards the end of the Millennium Drought, and should serve to only be broadly indicative of the regional geology. At a regional scale, all inversions do indeed produce a consistent geoelectric section in Figure~\ref{pic:deterministic}. {Differences between the various inversions both in terms of inverted conductivity and RMS values achieved are presented in Figure~\ref{pic:deterministic_rms}}.
\subsection{Probabilistic inversion results}
\begin{figure}
	\centering
	\includegraphics[width=\linewidth]{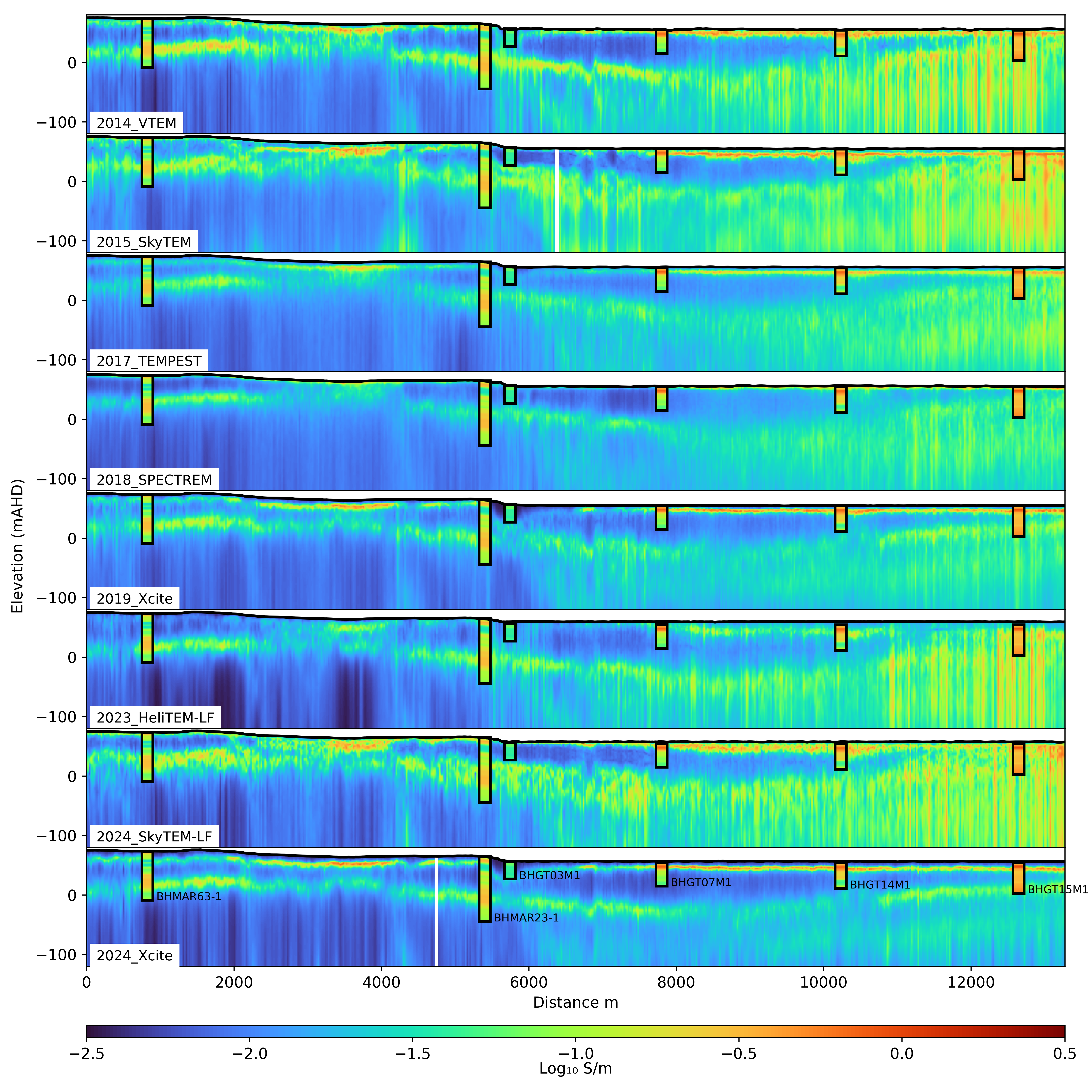}
	\caption{Sections showing the 10th posterior percentile of conductivity with depth from Bayesian inversions of the same AEM data used in Figure~\ref{pic:deterministic}. This percentile is representative of the resistive end of probabilistically inverted conductivities. If this percentile is \textit{conductive} (red) at a given location, there is a high probability that the subsurface at that location is conductive. As this percentile is representative of the resistive end of the conductivity spectrum, conductivities should generally be lower (i.e., deeper blue) than the downhole conductivities.}
	\label{pic:P10}
\end{figure}

\begin{figure}
	\centering
	\includegraphics[width=\linewidth]{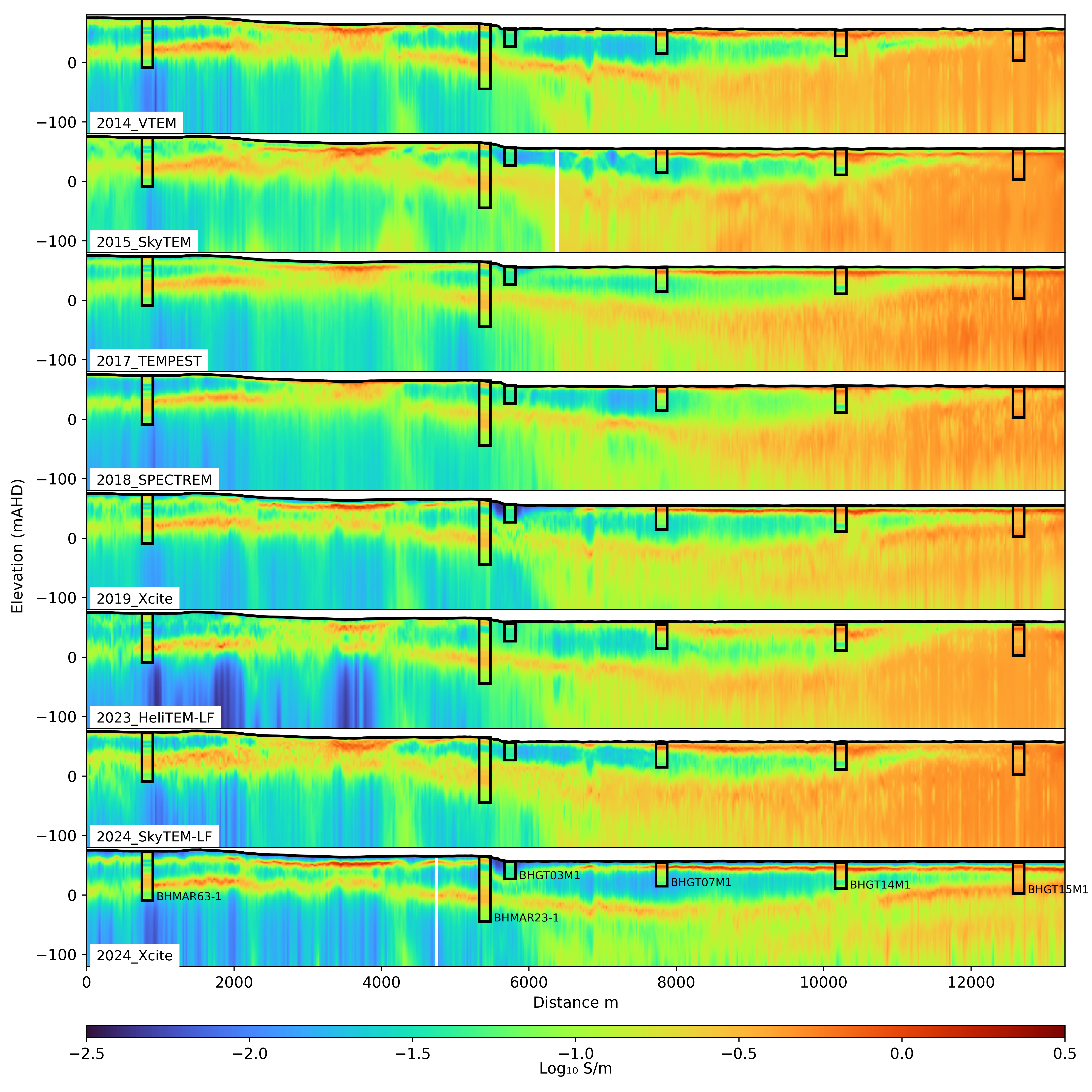}
	\caption{Sections showing the 50th posterior percentile or median conductivity with depth from Bayesian inversions of the same AEM data used in Figure~\ref{pic:deterministic}. This percentile is representative of an ``overall'' conductivity of the subsurface. Values should generally be similar to downhole conductivities. The median conductivities are similar to the deterministic conductivities in Figure~\ref{pic:deterministic} with the exception that the Bayesian inversion models are not regularised to return to a background resistive reference value.}
	\label{pic:P50}
\end{figure}

\begin{figure}
	\centering
	\includegraphics[width=\linewidth]{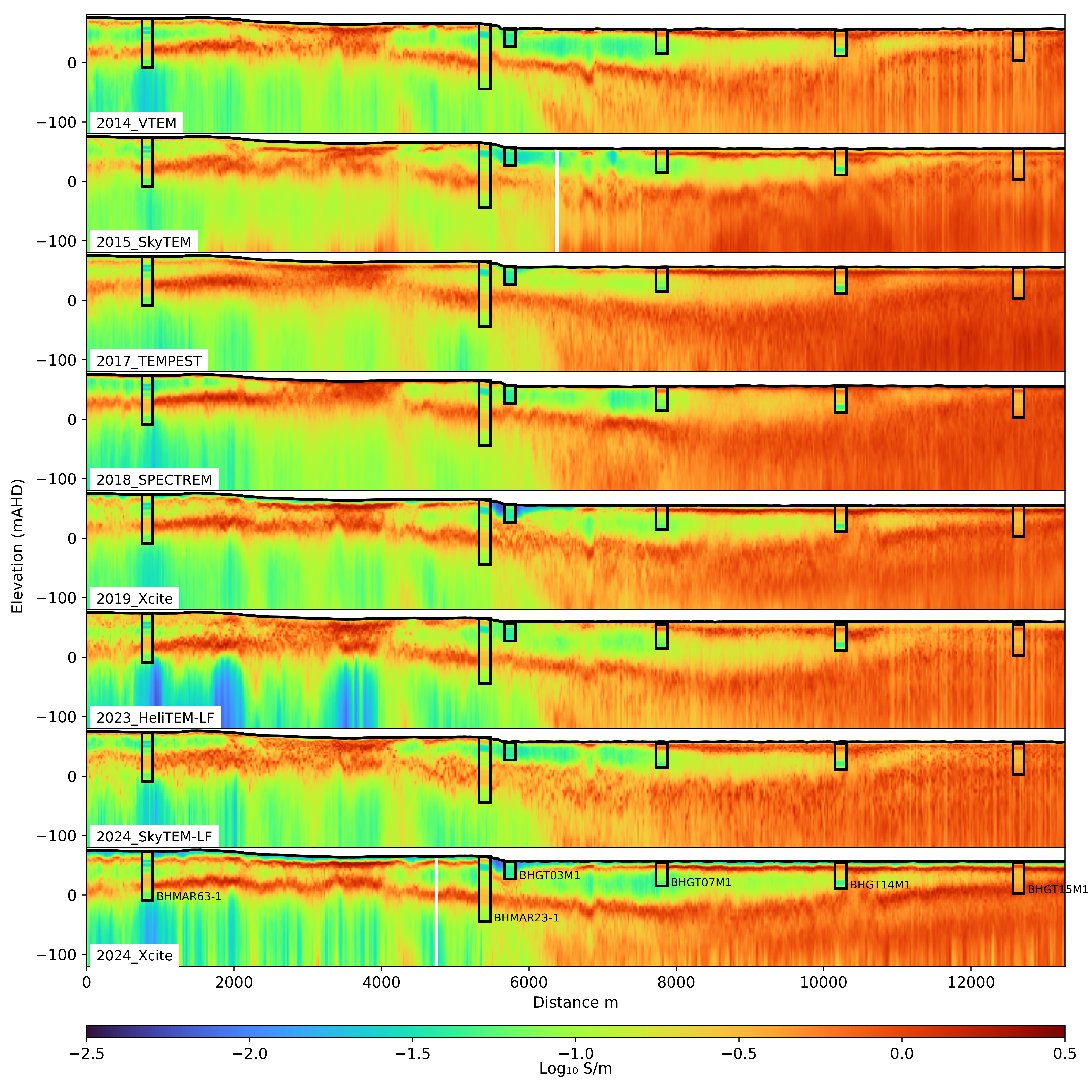}
	\caption{Sections showing the 90th posterior percentile of conductivity from Bayesian inversions of the same AEM data used in Figure~\ref{pic:deterministic}. This percentile is representative of the conductive end of probabilistically inverted conductivities. If this percentile is \textit{resistive} (blue) at a given location, then there is a high probability that the subsurface at that location is resistive. As this percentile is representative of the conductive end of the conductivity spectrum, conductivities should in general be higher (i.e., warmer red) than the downhole conductivities.}
	\label{pic:P90}
\end{figure}

As with the deterministic inversions, the fixed-wing systems were inverted using the total amplitude of the magnetic field in the receiver X-Z plane. Sections constructed from the marginal posterior inverted conductivities showing the 10th, 50th (median) and 90th percentiles are shown in Figures~\ref{pic:P10},~\ref{pic:P50} and~\ref{pic:P90}, again with RMS data misfits higher than 1.5 blanked out. In general, the logged induction downhole conductivities are observed to fall in-between the inferred 10th and 90th percentiles, despite the amount of time that has elapsed between the borehole logging and the AEM acquisition. For any system, a low spread of conductivity between different percentiles indicates low posterior uncertainty. Unlike the deterministic inversions, the probabilistic inversions are not regularised to be smooth, nor are they regularised to trend to $\mathbf m_0$ when losing sensitivity to data. With the probabilistic percentiles, it is now possible to infer that the western part of the section (2 km along line) is indeed resistive below 0 m height, while between 8 to 10 km along the line, we encounter from the surface downwards, a conductive-resistive-conductive sequence which is highly uncertain beneath -50 m. 

\begin{figure}
	\centering
	\includegraphics[width=0.7\linewidth]{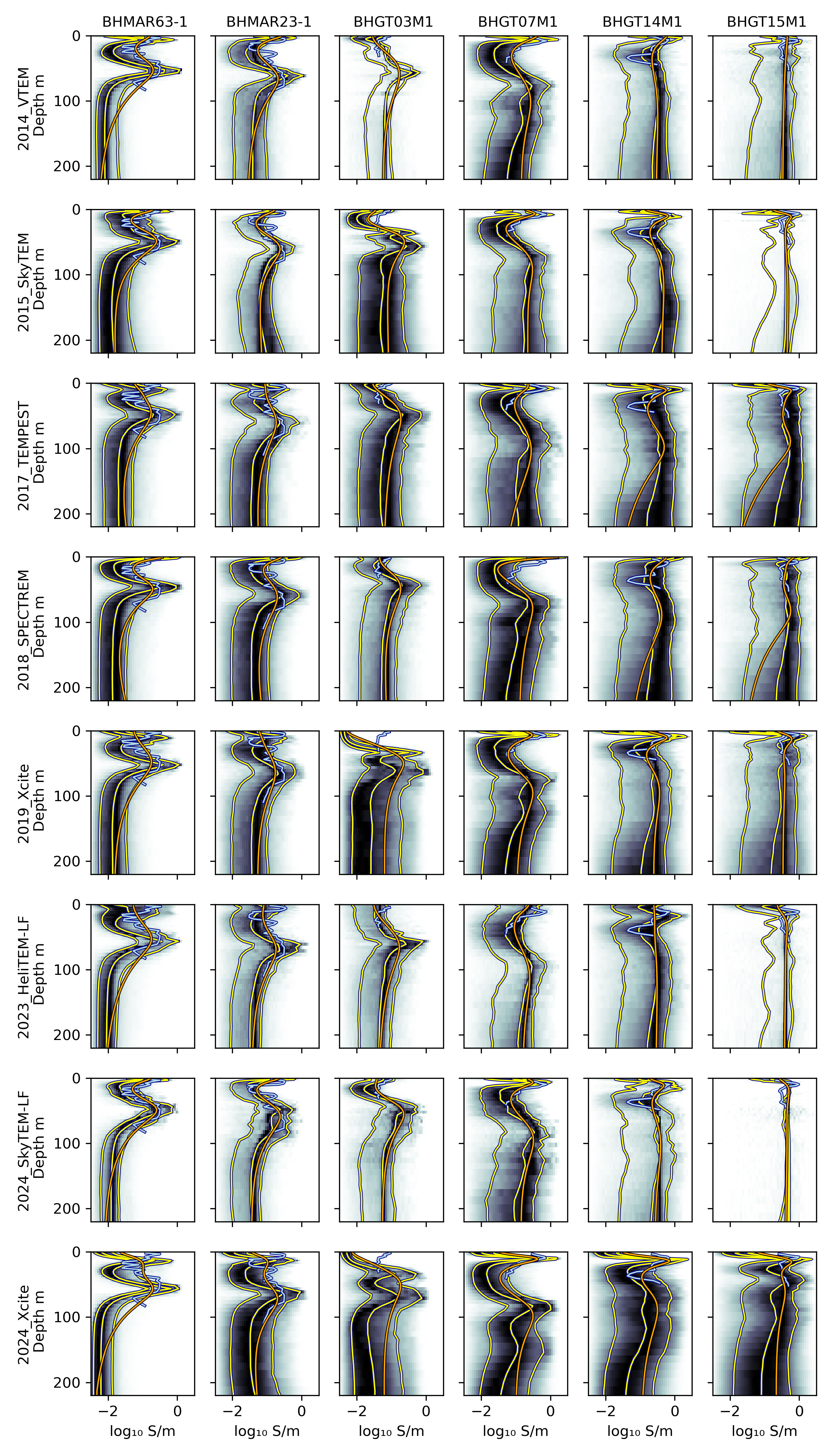}
	\caption{Greyscale images of posterior marginal probability density functions of conductivity with depth shown at borehole locations, arranged chronologically (top-bottom) and northwest to southeast (left-right). Each column represents a particular borehole location (refer to Figure~\ref{pic:map}), and each row, an AEM survey. Darker shades indicate higher probability. Each depth has been normalised to show a maximum probability density of 1. Also overlain, are downhole conductivities logged in each borehole (blue lines), and the 10th , 50th and 90th percentiles of conductivity (yellow lines) at the borehole locations in Figures~\ref{pic:P10},~\ref{pic:P50} and~\ref{pic:P90}. The deterministic inversion result from Figure~\ref{pic:deterministic} is indicated by an orange line. For the most part the downhole conductivities and the deterministic inversion lie within the posterior credible region between the 10th and 90th percentiles.}
	\label{pic:wells_post_depth_deep}
\end{figure}

Posterior marginal probability densities of conductivity with depth at the well locations are shown as images in Figure~\ref{pic:wells_post_depth_deep}, with darker shading indicating higher posterior certainty. The 10th, 50th, and 90th percentiles are overlain, as are the logged induction conductivities. The deterministic inversion result at the same location has also been plotted. As observed on the percentile sections, with the exception of the Xcite system at shallow depths at BHMAR23-1, all systems generally keep the downhole logged conductivities between the extremal 10th and 90th posterior conductivity percentiles. We cannot be certain that the shallow Xcite results at that location are an artefact. This is because the Xcite system forward modelling requires the convolution of a system response during transmitter on-times (very early times) that theoretically could map the near surface accurately. However, as mentioned earlier, any inaccuracies in the system response filter will propagate into the inversion result. Another feature of note from Figure~\ref{pic:wells_post_depth_deep} is that the low base frequency systems (HeliTEM-LF 2021 and SkyTEM-LF 2024), which theoretically should provide deeper signal penetration, do indicate remarkably low posterior uncertainty even at depths beyond 200 m, despite providing smoother (low frequency) inversion images. 
\section{Interpretation of the inversion results}
\subsection{Picking geological boundaries}
\begin{figure}
	\centering
	\includegraphics[width=\linewidth]{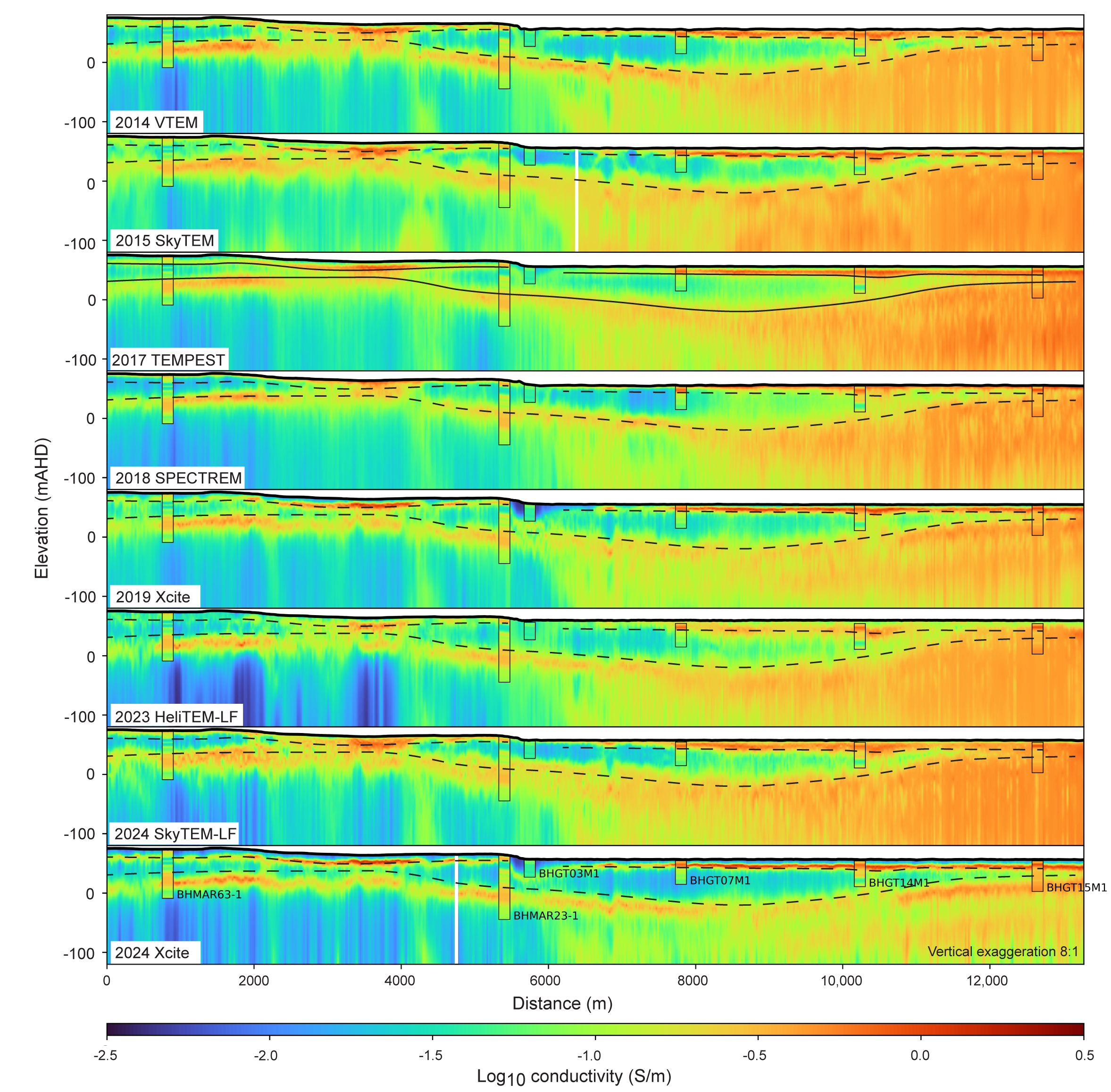}
	\caption{Sections showing the 50th posterior percentile of conductivity (median) overlain with lines highlighting major conductivity contrasts. Interpretation based on the 2017 TEMPEST data (solid lines in 3rd panel) is overlain on all other inverted sections (dashed lines). Note the subtle differences in the locations of features between the inverted surveys.}
	\label{pic:AEM_sections}
\end{figure}

An excellent heuristic for finding conductors using probabilistic inversions is to look for conductive areas in the 10th posterior percentile of conductivity. If the 10th percentile is conductive (Figure~\ref{pic:P10}), 90\% of conductivities at that depth location are conductive. Conversely, high probability resistive areas are represented by low conductivity within the 90th posterior percentile of conductivity (Figure~\ref{pic:P90}). In conjunction with the borehole geophysical data, and noting that interfaces between geological layers are often indicated by changes in conductivity with depth, we turn our attention to just the posterior conductivities from the 2017 Tempest survey. Interfaces were manually interpreted (picked) for this section where conductivity changes with depth (as opposed to at the peak of conductivity) and are shown as solid lines (Figure~\ref{pic:AEM_sections}). The \textit{same} picks from the 2017 TEMPEST data are overlain on all sections as shown in Figure~\ref{pic:AEM_sections} as dashed lines. {We emphasise here that there is no particular reason behind the choice of the 2017 TEMPEST results for the reference interpretation. The interpretation based on it is simply used for examining how it appears on sections made from the other AEM systems. While we have investigated automated picking tools \cite[e.g.,][]{Wilford2025AEMAssist}, our experience in interpreting data from across nearly two-thirds of the Australian continent \citep[e.g.,][]{Wong2023MultilayeredChronostratigraphic} at 20-km line spacing is that great care needs to be applied in the used of automated picking and we are not yet confident of its use across different systems or variable geology.}

\begin{figure}
	\centering
	\includegraphics[width=0.85\linewidth]{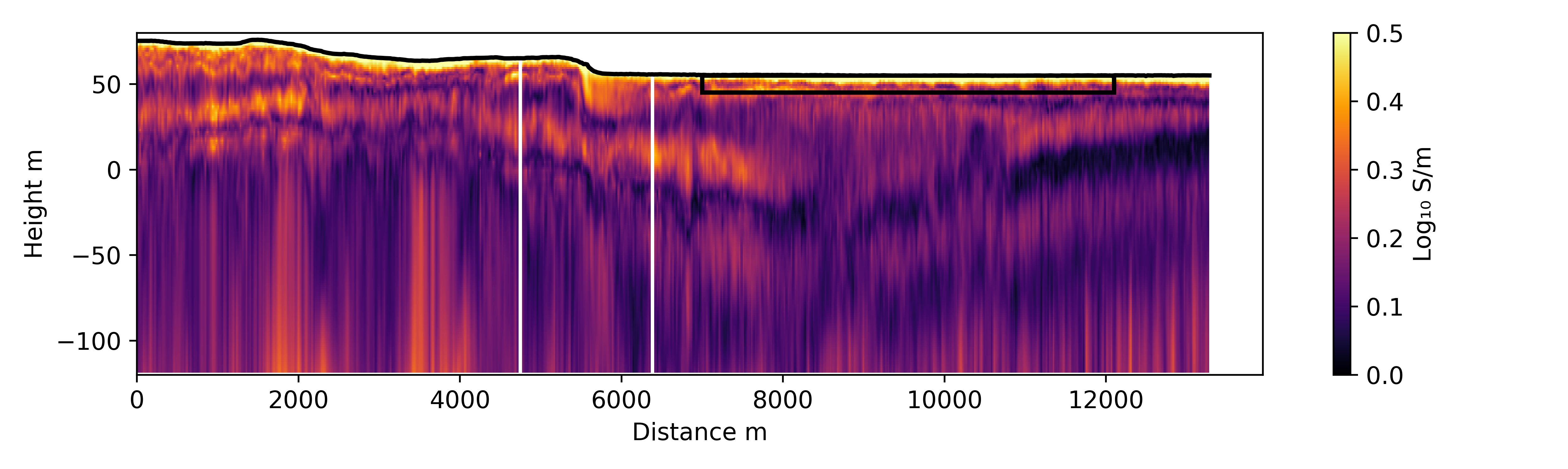}
	\includegraphics[width=0.85\linewidth]{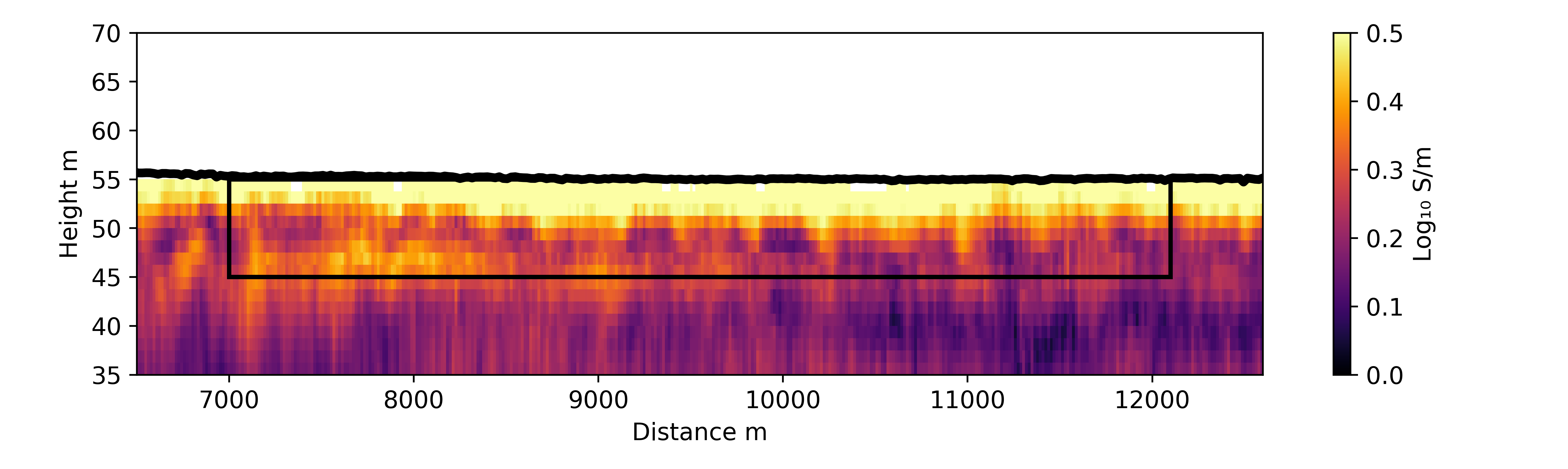}
	\caption{Visualisation of the pixel-wise standard deviation between all the median conductivity sections seen in Figure~\ref{pic:AEM_sections}. The greater the variability, the higher the standard deviation (incandescent colours). Variability is greatest where inverted conductivity values are different, or have shifted between the different inverted percentiles. The greatest variability is found in association with the lake, and is indicated with a 10 m deep, 5 km long box in the upper panel. The bottom panel shows detail within the lakebed area, with the boxed area used to calculate an average ``bulk" conductivity in each percentile for a regression analysis.}
	\label{pic:sd_median}
\end{figure}

We observe subtle shifts in the positions of conductors between the sections, noting that vertical exaggeration is about $\sim$8:1. To examine this more quantitatively, we assess the standard deviation between the 8 inverted median conductivity section images in Figure~\ref{pic:AEM_sections} at every pixel location. The resulting image of standard deviations in the median conductivity is shown in Figure~\ref{pic:sd_median}. The highest standard deviations are found in the top 10 m or so, noting that the regions of high conductivity variability across the median section are found both over the lake as well as onshore to the west. 
\subsection{Establishing a relationship between water volume and lakebed conductivity}
\begin{figure}
	\centering
	\includegraphics[width=1\linewidth]{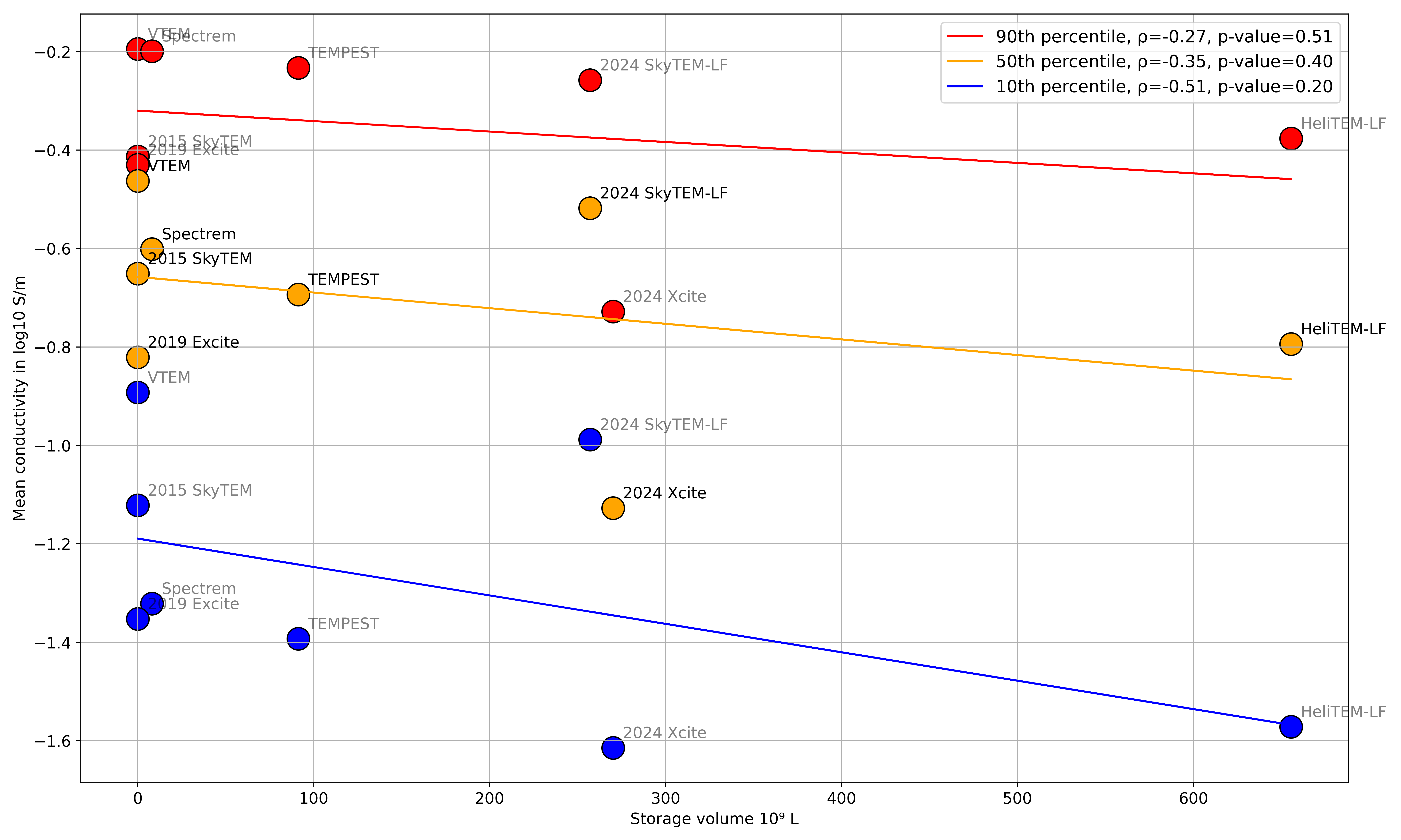}
	\caption{Best fit lines found between the bulk percentile conductivities in the box  marked in Figure~\ref{pic:sd_median} and Lake Menindee water volumes shown in Figure~\ref{pic:waterlevels}. The most representative relationship is between the bulk median conductivity and the water volumes shown in orange. The 10th percentile (blue) and 90th percentile (red) relationships are also plotted. The correlations, together with their significance are shown in the legend. The predetermined value of 0.05 for a significant relationship to hold, cannot be established.}
	\label{pic:correlation}
\end{figure}

Since it is the lake which experiences water level variability, we zoom in on a zone 5 km long and 10 m deep. This zone encompasses the conductivity variability (Figure~\ref{pic:sd_median}) beneath the dry lakebed surface, and we look here for a linear relationship and associated correlation between the water volume in the reservoir and the bulk median conductivity inferred by our surveys over time (Figure~\ref{pic:correlation}). Since the 10th and 90th percentiles highlight high conductivity and low conductivity respectively, best fit relationships and correlations have been found for these percentiles as well. It can be argued that evaporative concentration of salts during drier periods provide a source of salt that can be progressively mobilised downwards into the regolith and shallow groundwater beneath the lake bed (leading to higher conductivity, depending on the underlying water saturation). However, the Cenozoic lakebed sediments in the southeast are clayey and may not allow permeable flows of ions away from the boxed zone in Figure~\ref{pic:sd_median} (and hence lower conductivity), even with inflows of water into the reservoir. Since we know there can be a change in conductivity with reservoir water levels, but lack exact knowledge of the factors that determine the direction of change, we carry out a standard two tailed $t$-test with a prior significance value of 0.05. For 8 observations using the median conductivities, the correlation $\rho=- 0.35$, the $p$ value obtained is 0.40. This is far greater than the required standard of $p<0.05$ for significance. 

None of the percentiles show a significant correlation. Details of how correlation, the best fit line, and significance are related, computed and used, is given in Appendix~\ref{A:corrsig}. {Additionally, we propagate McMC posterior conductivity uncertainty to the water volume and conductivity relationship (Appendix~\ref{sec:McMC_prop}). We find that only a small minority (0.2\%)of $10^6$ sampled correlations show significant correlation $\rho <-0.71$, with the median correlation similar to what is reported in Figure~\ref{pic:correlation}. Though we do see a moderate negative correlation, we cannot robustly establish a significant correlation between the water volumes in the reservoir and conductivities in the top 10 m of the lake underneath the dry lakebed surface.}
\subsection{Geological cross section}
\begin{figure}
	\centering
	\includegraphics[width=1\linewidth]{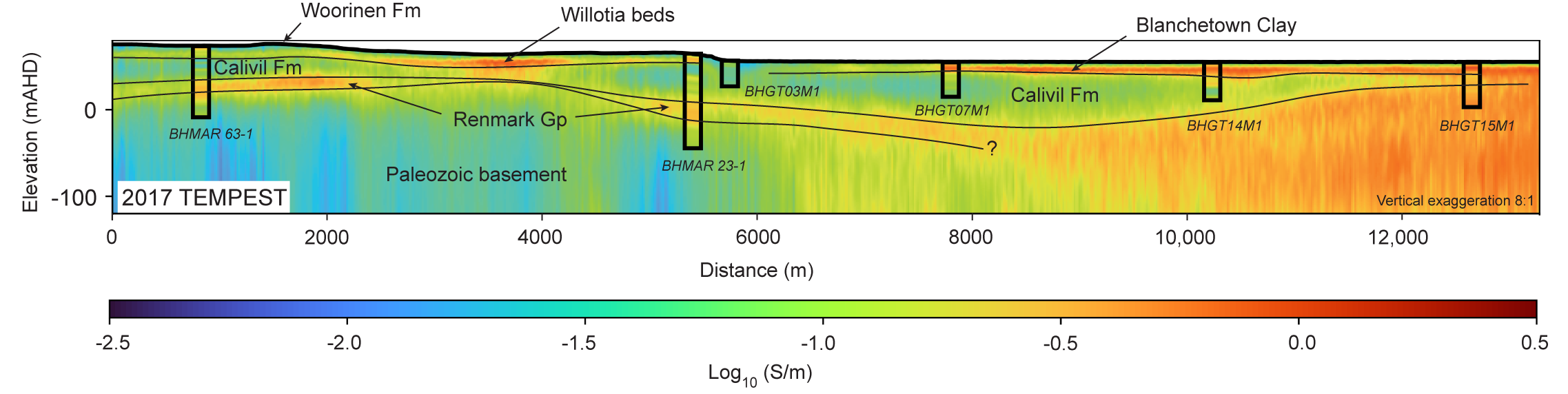}
	\caption{Conductivity-depth section (median) with interpreted geological formations based on a combination of available stratigraphic records, borehole induction log conductivity values, and observations of outcropping geology \citep{BHMAR_report}. Strong geoelectrical contrasts clearly identify the top and bottom of the relatively resistive Calivil Formation (Fm) and define the base of the conductive Renmark Group (Gp) across much of the section, while shallow clay-rich facies of the Willotia beds and the Blanchetown Clay are also imaged as discrete, thin conductors.}
	\label{pic:geology_sections}
\end{figure}

Despite being unable to establish a relationship between lake water volumes and shallow conductivity in Menindee Lake, the available stratigraphic information and accompanying borehole geophysics (induction conductivity) \citep{BHMAR_App, BHMAR_report, Rockprops2019} confirm that the models permit clear identification of key Cenozoic stratigraphic units of the Murray Basin \citep{Murraygeol} (Figure~\ref{pic:geology_sections}).

The very shallowest (mainly resistive) parts of the section are weathered red-brown calcareous and clayey sands of the Woorinen Formation, which comprises closely spaced longitudinal dunes ranging in length from 0.5 to 3 km and typically 2–6 m high \citep{Mallee}. Along the western part of the line the underlying Willotia beds represent 7–10 m thick ancient floodplain deposits consisting of a fine-grained muddy facies. These sediments were laid down in dominantly fluvial overbank settings with minor lacustrine deposition \citep{BHMAR_report}, and facies variation produces differing geo-electrical responses in the shallow section (Figure~\ref{pic:geology_sections}).

The dominant shallow, thin conductor in the east (Figure~\ref{pic:geology_sections}) is consistent with the presence of the Blanchetown Clay, a green-grey and red to brown clay unit $<$ 10 m thick \citep{Murraygeol} associated with lacustrine deposition of paleo-Lake Bungunnia \citep[e.g.,][]{Blanchetown}.

Across most of the section these shallower Late Cenozoic units overlie the Late Miocene to Early Pliocene Calivil Formation -- a poorly consolidated pale grey, poorly sorted coarse to granular quartz sand with a white kaolinitic matrix (Figure~\ref{pic:geology_sections}). The unit is dominantly fluvial in origin, forms an extensive blanket up to 100 m thick (usually $<$ 60 m thick) beneath much of the riverine plain, and represents one of the main aquifer units within the Murray Basin \citep{Murraygeol}. Where groundwater within the unit is of relatively low salinity the geoelectric response of the unit is most often resistive.

Lying unconformably beneath the Calivil Formation are the Eocene to Late Miocene upper sequences of the Renmark Group (Figure~\ref{pic:geology_sections}). The unit comprises unconsolidated to poorly consolidated, thinly bedded dark brown to grey sand and clay, with occurrences of black carbonaceous silt, brown coal and peat. The unit is commonly micaceous, pyritic and ferruginised, and includes intercalations of poorly sorted fine to medium quartz sand, polymictic sand, and sandy dolomite \citep{Murraygeol}. The unit is reported as being up to 200 m thick, but in areas of shallow basement, such as the Menindee region, the unit may be significantly thinner owing to erosion or reduced initial deposition. The nature of the materials comprising the Renmark Group and the observation that it commonly contains saline groundwater \citep[e.g.,][]{Macca2024} are consistent with its dominantly conductive response.
\section{Quantitative measures of information, and caveats on their uses}
\begin{figure}
	\centering
	\includegraphics[width=\linewidth]{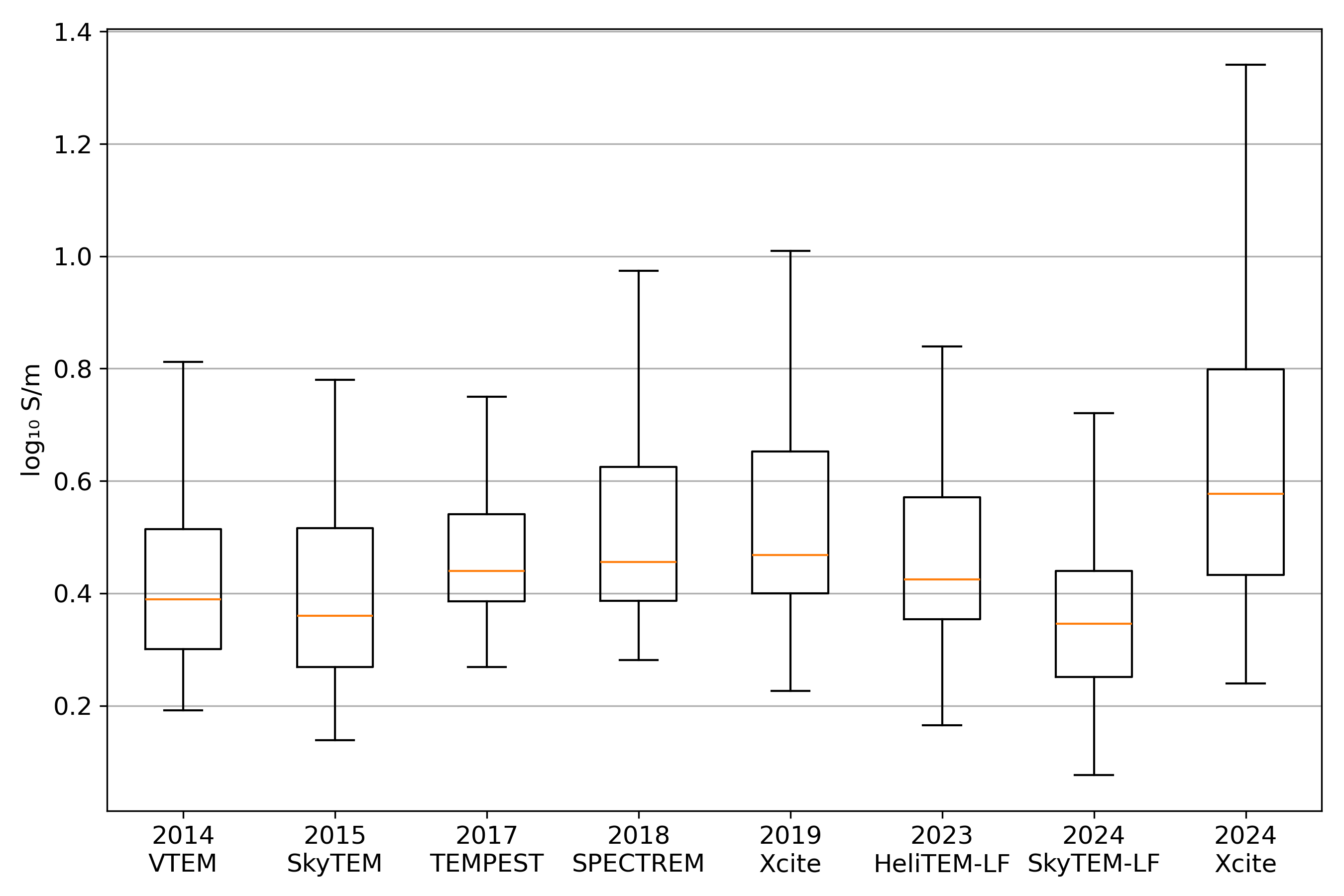}
	\caption{Box and whiskers plot of mean absolute ``error'' (MAE) or deviation of posterior McMC samples from borehole log observations, for all systems, across all boreholes. Orange represents the median MAE, box represents the interquartile range with 50\% of the probability mass.}
	\label{pic:MAE_boxplot}
\end{figure}
\subsection{Mean absolute error (MAE)}
As the posterior conductivities with depth are not Gaussian in Figure~\ref{pic:wells_post_depth_deep}, we adapted the following ensemble spread \cite[e.g.,][]{Nester2012, Jha2018} measure of closeness to the borehole induction log data:
\begin{equation}
\text{MAE}(z) = \frac{1}{N_\text{McMC}}\sum_{i=1}^{N_\text{McMC}} |\sigma_{\text{inv}_i}(z) - \sigma_{\text{log}}(z)|, 
\end{equation}
where $|\cdot|$ is the absolute value operator, $\sigma_{\text{inv}_i}(z)$ is the $i^{th}$ McMC sampled conductivity at $z$, $N_\text{McMC}$ is the total number of samples in the posterior ensemble, and $\sigma_{\text{log}}(z)$ is the induction log value at the same depth. These calculations were carried out at all logged depths, across all boreholes, for each system, and the results are plotted in Figure~\ref{pic:MAE_boxplot}. Lower MAE values and a tighter spread indicate closer proximity to the observed borehole induction log value across all depths. Since the boreholes were logged 4 years prior to any survey flight and the effect of water saturation in the system is difficult to determine exactly, closeness to zero in these values should be treated with caution. 
\subsection{Information gain}
Another well studied and rigorous measure of information contained within a probability density function (PDF), is entropy \citep{Lindley1956}. Relative entropy, or information gain between a prior and posterior PDF \citep{Renyi1961}, is given by the Kullback-Leibler divergence \citep{kld} between the prior and posterior PDFs. The utility of Kullback-Leibler divergences have been extensively documented in the context of AEM inversion in Section 4 of \cite{Ray2023}, and can be summarised as follows. Zero Kullback-Leibler divergence between prior and posterior indicates no information gain after carrying out the experiment, which in our case, is an AEM survey and subsequent inversion of conductivity. Higher information gain corresponds to narrower posterior conductivity distributions, and hence, more information gain. The advantage of using information gain is that it can be computed at all depths, and does not depend on having a reference observation such as a borehole log, which does not cover all areas and depths of interest. We calculate the information gain at all borehole locations, using the density ratio estimation method of \cite{Sugiyama2013} and plot the results in units of bits of information, for three different depth ranges $z$ (Figure~\ref{pic:KLD_boxplot}). A brief mathematical overview of information gain is given in Appendix~\ref{sec:KLD}.

While calculations of information gain are indeed comparable, even in areas and depths with no borehole information -- they are dependent on the subsurface conductivity being imaged, which could change over time.
\begin{figure}
	\centering
	\includegraphics[width=\linewidth]{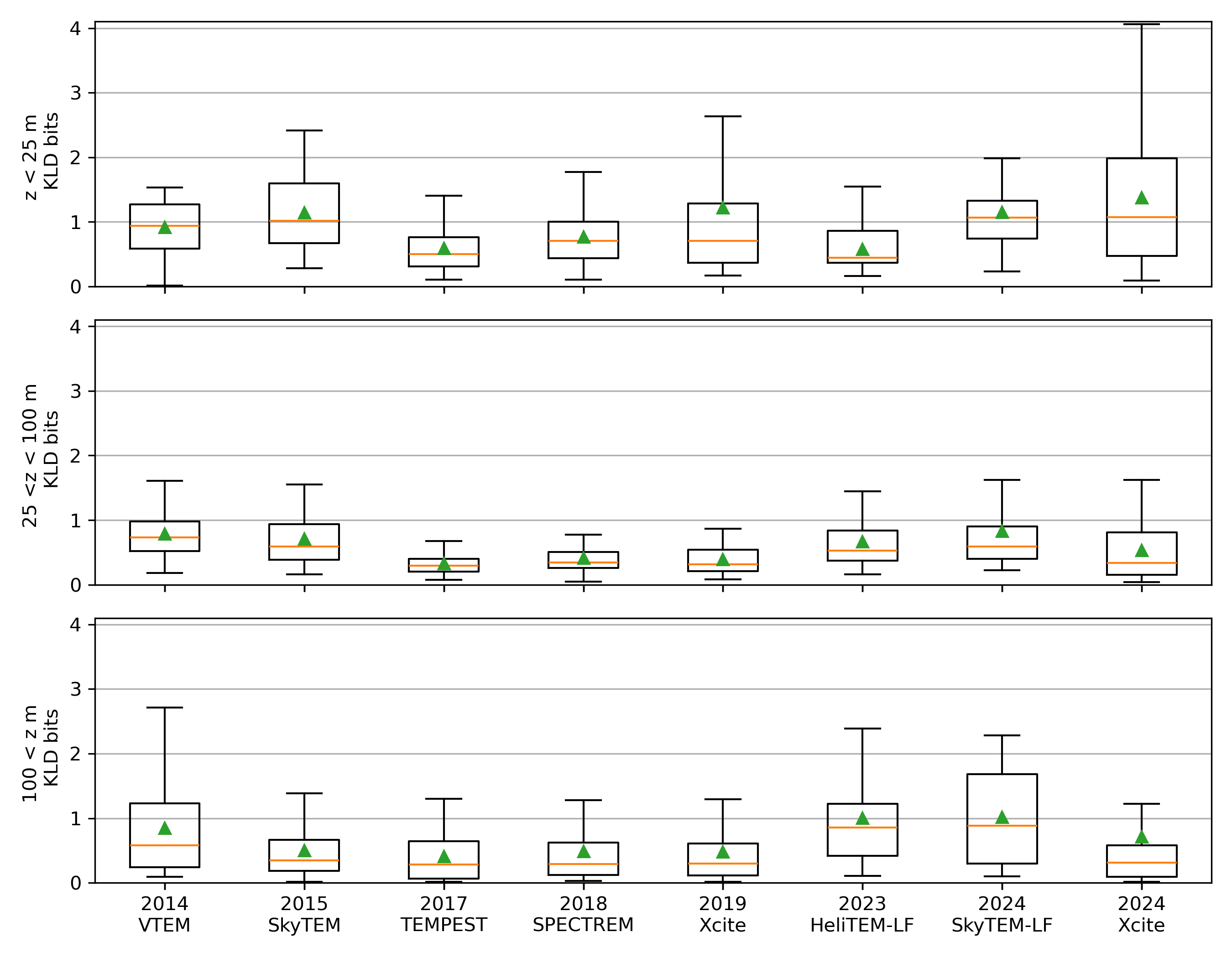}
	\caption{Box and whiskers plots representing the information gained by each AEM survey, given by the Kullback-Leibler divergence (KLD) from the prior to the posterior.  Plots have been made for shallow (top row), medium (middle row) and deeper depths (bottom row). Green triangle is the mean KLD, representative of the experimental information gain of the surveying system.}
	\label{pic:KLD_boxplot}
\end{figure}
\section{Discussion}
In this work, we have compared a decade of AEM overflying an area of reasonably well known geology encompassing a surface water reservoir experiencing large ranges of flux through time. We have inverted the survey data using both deterministic and Bayesian probabilistic methods. Care has been taken to account for the nuisance parameters in fixed-wing systems to minimise their impact on inverted subsurface conductivity. While a deterministic Occam-type inversion provides an excellent first-pass representation of the geologically influenced conductivity variation, model regularisation is necessary to provide a smooth and stable solution. This often removes \textit{a posteriori} high probability features in a single deterministic solution. 

Probabilistic inversions, on the other hand, get around these smoothness constraints by requiring minimal prior information on parameter and nuisance values, and produce an \textit{a posteriori} ensemble of solution models. It must be stressed that any one model is not interpretable by itself. However, the posterior ensemble of the geological parameter of interest, i.e., conductivity, varies smoothly with depth except where the data and prior require all members of the ensemble to exhibit sharp changes. While looking at thousands of posterior ensembles and their distribution is difficult to do and is an active area of research, median and end-member percentiles (10th and 90th) of conductivity provide valuable insights into the the nature of subsurface conductivity variation, and the confidence with which we can interpret it. 

Bayesian uncertainty only propagates rigorously through the parameters we have successfully set priors on, namely transmitter-receiver nuisances (for fixed-wing systems) and subsurface conductivities. There are various aspects of the AEM surveys, both identified and unknown, that we have not accounted for. Some of these are inaccuracies in the transmitter waveform and modelling physics, incomplete or overzealous primary field removal, insufficient knowledge of receiver characteristics, and noise models used. While noise in the data are measured and statistically assigned from repeat lines \citep{Green2003}, for the other factors, inaccuracies will either propagate into inverted models, or prevent us from teasing out all the information we would like to obtain about the shallowest depths. 

Small-scale differences in inferred conductivity between different surveys is potentially due to changes in subsurface conductivity as well as changes in the AEM system (or if the same system, changes in the specifics about the AEM system and its hardware), which we are unable to separate. For instance, the large standard deviations at 2 km and 4 km line distance, from 0 m downwards in the top row of Figure~\ref{pic:sd_median} are in all likelihood due to the HeliTEM-LF system providing a clear image of resistivity at depth (Figure~\ref{pic:AEM_sections}), given the low base-frequency operation of that system -- not because there is some change in the subsurface geoelectric character over time. For system nuisances, one approach to marginalise these could be to use machine learning methods applied hierarchically as proposed by \cite{Morikawa2023} for observational astronomy. 
\section{Conclusions}
AEM can indeed be used to map changes in subsurface conductivity such as salt water intrusions \citep[e.g.,][]{Gottschalk2020}. True time lapse inversion as detailed by the methodology in \cite{Doetsch2010} {using ERT and GPR} to invert for an \textit{updated} conductivity model (as opposed to independently inverted) can perhaps tease out subtle changes using AEM surveys, and more work is needed on this front. Though it is well known, we reiterate that AEM sensitivity is not the same throughout the conductivity range of interest. A perturbation of conductivity at the resistive end (e.g., 0.01 S/m) produces a much smaller change in the AEM response than at the conducting end (e.g., 1 S/m) -- which is one reason for the large spread of conductivity values in the resistive 10th posterior percentile.

From the investigations presented here, we advocate caution when interpreting subtle changes in conductivity over time using repeated flyovers of the same location and independent inversions of conductivity from each pass. On the positive side, regardless of the AEM system used, we are able to provide a consistent image of the regional geology, independently validated by a combination of known stratigraphy and borehole induction log data. This serves to strengthen our view that AEM imaging is a robust technology that can be used for far more than ``anomaly hunting'', lending itself to a wide range of geoscience applications. 
\section*{Acknowledgments}
This work would not have been possible without close cooperation from all the AEM vendors who have flown our test range. Discussions with our fellow geoscientists over many years, particularly at the ASEG Discover meeting in Tasmania in November 2024 were particularly insightful. This study was funded by the Australian Government's Resourcing Australia's Prosperity initiative, with a view to increasing precompetitive geoscience knowledge. All computation was carried out on the \textit{Gadi} high performance computing cluster at the National Computational Infrastructure, Australia, at the Australian National University, Canberra, Australia. We thank David Lescinsky for arranging our computer time competitively between Geoscience Australia's various projects. 

All computation was carried out using the open source Julia language \citep{Bezanson2017}. HiQGA v0.4.13 is available at \url{https://github.com/GeoscienceAustralia/HiQGA.jl/releases/tag/v0.4.13}.

{We thank two anonymous reviewers for their constructive comments, particularly around Monte Carlo uncertainty propagation into the water-volume conductivity relationship.}

\section*{Data Availability}
Test range data are provided on a commercial in confidence basis and cannot readily be shared -- however, we use and modify the open-source HiQGA code base to perform the geophysical inversions. The code base with example notebooks and data, some of which is from the test range, can be freely cloned from \url{https://github.com/GeoscienceAustralia/HiQGA.jl}.

\bibliographystyle{plainnat}
\bibliography{library}
\appendix
\section{Relationship between the least squares fit and the correlation coefficient} \label{A:corrsig}
When regressing observations $y_i$ dependent on changes in a quantity $x_i$, linear regression has us proceed such that we can write:
\begin{align}
y_i &= c_1x_i + c_2.\label{eqn:linreg}\\
\intertext{The coefficients $c_1$ and $c_2$ can be found using ordinary least squares and the following objective function $\phi(c_1, c_2)$}
\phi(c_1, c_2) = &\sum_i(y_i - c_1x_i - c_2)^2.
\intertext{To minimise the above objective function with respect to $c_1$ and $c_2$, we set both $\pdv{\phi}{c_1} = 0$ and $\pdv {\phi}{c_2} = 0$ to obtain:}
c_1 &= \frac {\Cov(x,y)}{\Var(x)},\label{eqn:c1}\\
\intertext{and}
c_2 &= \E(y) - c_1\E(x), \label{eqn:c2}
\end{align} 
where $\Cov$ is covariance, $\Var$ is variance and $\E$ is the expectation. We know that the correlation $\rho(x,y)$ between $x$ and $y$ is defined as:
\begin{align}
\rho(x,y) &= \frac {\Cov(x,y)}{\sqrt{\Var(x)}\sqrt{\Var(y)}},\\
	&= \frac {\Cov(x,y)}{\Var(x)}\frac{\sqrt{\Var(x)}}{\sqrt{\Var(y)}},\\
	\intertext{using \eqref{eqn:c1} we get}
\rho(x,y)  &= c_1\frac{\sqrt{\Var(x)}}{\sqrt{\Var(y)}}. \label{eqn:corr}
\end{align}
Therefore, the relationship between the line of best fit for $y$ against $x$ and the correlation $\rho(x,y) $ is established by equations \eqref{eqn:linreg}, \eqref{eqn:c1}, \eqref{eqn:c2} and \eqref{eqn:corr}. 

It is not enough to establish a correlation between variables with a small number of observations to comment on their significance, for which a $t$-test with a preset significant $p$ value such as 0.05 is used. The $t$ statistic is computed as follows \citep{Field2012}:
\begin{equation}
	t = \rho\frac{\sqrt{N-2}}{\sqrt{1-\rho^2}}, \label{eqn:tstat}
\end{equation}
where $N$ is the number of observations. The null hypothesis that the correlation coefficient is zero can be rejected if  the the $t$-test for $t$, $\rho$ and $N$ in \eqref{eqn:tstat} provides a $p$ value smaller than the preset 0.05, i.e., the established relationship is then deemed to be significant.
\section{{Monte Carlo propagation of AEM posterior uncertainty into water volume--conductivity relationship}}\label{sec:McMC_prop}
\begin{figure}
	\centering
	\includegraphics[width=\linewidth]{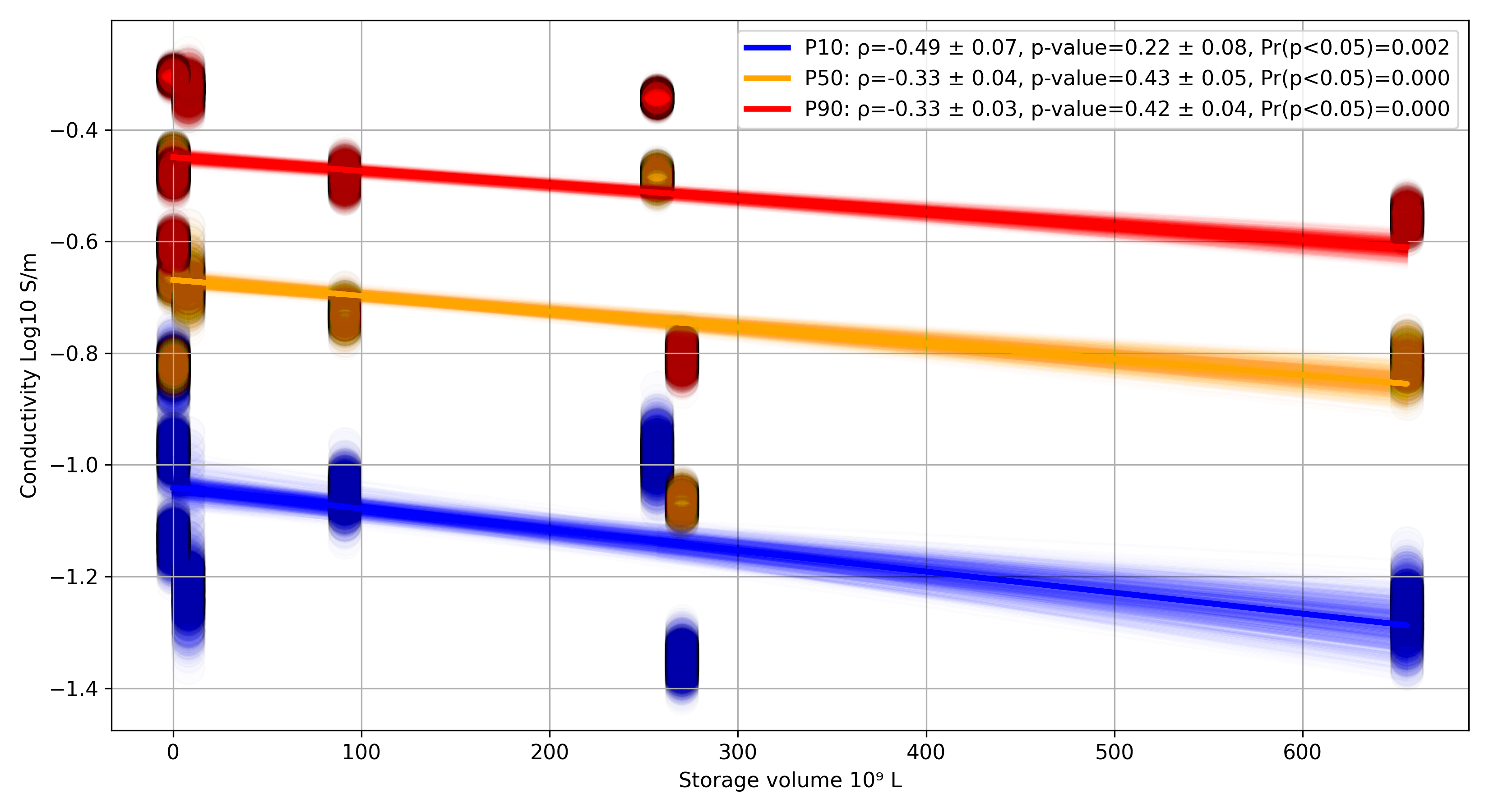}
	\caption{{Realisations from the McMC propagation of conductivity uncertainty to correlation with lake water volume, plotted with transparency. Only 0.2\% of 10th percentile conductivities in the boxed region in Figure~\ref{pic:sd_median} show significant correlation with $p<0.05$. The vast majority of realisations across all percentiles show moderate negative correlation not strong enough to meet the preset significance criteria.}}
	\label{pic:corr_mcmc}
\end{figure}
{The posterior conductivity uncertainty from the McMC inversion was propagated into the water volume--conductivity relationship as follows. We repeatedly sampled survey-level conductivity percentiles (10th, 50th and 90th) from the McMC inversion derived posterior conductivities in the boxed region highlighted in Figure~\ref{pic:sd_median} to form an 8-survey conductivity tuple. We then computed the resulting regression/correlation with the 8 water volumes and its two-sided p-value, repeating this procedure $10^6$ times. The results are shown in Figure~\ref{pic:corr_mcmc}. Only 0.2\% of computed regressions were found to have significant p-value (i.e., $p<0.05$), and only when looking at 10th percentile conductivity values. This analysis supports a consistent moderate negative correlation, while showing that the relationship is typically not strong enough to meet the significance criterion given the limited number of surveys.}
\section{The Kullback-Leibler divergence}\label{sec:KLD}
The Kullback-Leibler Divergence divergence in going from PDFs  $f$ to $g$ or $D_\text{KL}(g||f)$ (read ``$g$ with respect to $f$'', see \cite{Cover2005}) is given as:
\begin{align}
		D_\text{KL}(g||f) &= \int g(x) \log{\frac{g(x)}{f(x)}}dx,\\
		&=\E_{x\sim g(x)}\Big[ \log{\frac{g(x)}{f(x)}}\Big].
\intertext{Restricting ourselves to the univariate case without loss of generality, if the prior probability density is $p(m)$ and posterior density $p(m|d)$, then the information gain can be defined in terms of Kullback-Leibler divergence as follows:}
		D_\text{KL}\Big(p(m|d) || p(m)\Big) &= \int p(m|d) \log{\frac{p(m|d)}{p(m)}}dm, \\
		&=\E_{m\sim p(m|d)}\Big[\log{\frac{p(m|d)}{p(m)}}\Big]. \label{eqn:IG}\\ 
\intertext{Clearly, if the prior and posterior are equal, the log ratio equals zero, and no information is gained through the experiment. \cite{Lindley1956} says the ``average amount of information provided by the experiment" is provided by his equation 8, which is known as the Expected Information Gain or EIG \citep[see e.g., equation 10 of][]{Strutz2024}:}
		\text{EIG}(d, m) &= \E_{d\sim p(d)}\Big[\E_{m\sim p(m|d)}\Big[\log{\frac{p(m|d)}{p(m)}}\Big]\Big],\\
\intertext{which, using \eqref{eqn:IG} is the the expectation over data $d$ of the information gained in going from prior belief to posterior knowledge of the model $m$:}		
		\text{EIG}(d, m) &= \E_{d\sim p(d)}\Big[D_\text{KL}\Big(p(m|d) || p(m)\Big)\Big].
\end{align}
\end{document}